\documentclass[a4paper,12pt]{article}

\pdfoutput=1
\usepackage{amsmath}
\usepackage{amssymb}
\usepackage{amsfonts}
\usepackage{mathrsfs}
\usepackage{bbm}
\usepackage{graphicx,subfigure}

\usepackage{bookmark}

\usepackage{cite}

\usepackage{calc}
\usepackage{hyperref}
\usepackage{array}

\usepackage{ulem}

\usepackage{multirow}
\usepackage{color}
\usepackage{xcolor,colortbl}

\usepackage{psfrag}
\usepackage{pstricks}
\usepackage{epsfig}

\allowdisplaybreaks

\newlength{\dinwidth}
\newlength{\dinmargin}
\definecolor{nicered}{rgb}{1.0,0.0,0.2}

\definecolor{color1}{rgb}{0.9,.4,.2}

\definecolor{color2}{rgb}{0.3,.6,.7}

\definecolor{color3}{rgb}{0.7,.2,.7}

\usepackage{color}

\begin{document}

\title{
\vspace*{-0.5cm}
\bf \Large
Study of CP violation and CPT violation in $K^{\ast}(892)\rightarrow  K_{S,L}^{0}\pi$ decays at BESIII}

\author{Xiao-Dong Cheng$^{1}$\footnote{chengxd@mails.ccnu.edu.cn},  Ru-Min Wang$^{2}$\footnote{ruminwang@sina.com}, Xing-Bo Yuan$^{3}$\footnote{y@mail.ccnu.edu.cn}, Xin Zhang$^{4}$\footnote{xinzhang@hubu.edu.cn}\\
\\
{$^1$\small College of Physics and Electronic Engineering,}\\[-0.2cm]
{    \small Xinyang Normal University, Xinyang 464000, People's Republic of China}\\[-0.1cm]
{$^2$\small College of Physics and Communication Electronics,}\\[-0.2cm]
{    \small JiangXi Normal University, NanChang 330022, People's Republic of China}\\[-0.1cm]
{$^3$\small Institute of Particle Physics and Key Laboratory of Quark and Lepton Physics (MOE),}\\[-0.2cm]
{    \small Central China Normal University, Wuhan 430079, People's Republic of China}\\[-0.1cm]
{$^4$\small Faculty of Physics and Electronic Science,}\\[-0.2cm]
{    \small Hubei University, Wuhan 430062, People's Republic of China}\\[-0.1cm]}

\date{}
\maketitle
\bigskip\bigskip
\maketitle
\vspace{-1.2cm}

\begin{abstract}
{\noindent}The decays $K^{\ast}(892)\rightarrow  K_{S,L}^{0}\pi$ can be used to study CP violation and CPT violation. The $K^{\ast}(892)$ (hereinafter referred to as $K^{\ast}$) meson can be produced via $J/\psi$ decays at BESIII. In this paper, we study CP violation and $K_S^0-K_L^0$ asymmetry in the $J/\psi$ decays involving the $K^{\ast}$ meson in the final states and obtain the following results
\begin{align}
{\mathcal A}_{CP}\left(J/\psi\rightarrow K^{\ast}f_{J/\psi}\rightarrow K^0_{S}\pi f_{J/\psi}\right)=\left(3.64\pm0.04\right)\times 10^{-3},\nonumber\\
{\mathcal A}_{CP}\left(J/\psi\rightarrow K^{\ast}f_{J/\psi}\rightarrow K^0_{L}\pi f_{J/\psi}\right)=\left(-3.32\pm0.04\right)\times 10^{-3},\nonumber
\end{align}
and
\begin{align}
&R\left(J/\psi\rightarrow K^{\ast}f_{J/\psi}\rightarrow K^0_{S,L}\pi f_{J/\psi} \right)=\left(3.51\pm0.03\right)\times 10^{-3},\nonumber\\
&R\left(J/\psi\rightarrow \bar{K}^{\ast} \bar{f}_{J/\psi}\rightarrow K^0_{S,L}\pi \bar{f}_{J/\psi} \right)=\left(-3.45\pm0.03\right)\times 10^{-3}.\nonumber
\end{align}
Basing on two cases: the samples of $10^{10}$ and $10^{12}$ $J/\psi$ events, we calculate the expected numbers of the observed signal events on the CP violation and the $K_S^0-K_L^0$ asymmetry in the $J/\psi$ decays with $K^{\ast}$ meson in the final states, we find that the BESIII experiment may be able to unambiguously observe CP violation and $K_S^0-K_L^0$ asymmetry for each of these two cases. We study the possibility to constraint the CPT violation parameter $z$ and discuss the sensitivity for the measurement of $z$ in $J/\psi$ decays with $K^{\ast}$ meson in the final states at BESIII. The sensitivity of the measurement of $z$ depend on the measured precision of the parameters $m_L-m_S$, $\Gamma_L$, $\Gamma_S$, $p$ and $q$ and the consistence between the values of $t_0$ and $t_1$ and the event selection criteria in experiment.
\end{abstract}
\newpage

CP violation and CPT violation play important roles in deeping the understanding of the nature and studying physics beyond the Standard Model~\cite{Sakharov:1967dj,Luders:1957bpq,Riotto:1998bt,KLOE-2:2018yif,DiDomenico:2020pxd}. The decays with neutral K meson in the final states can be used to study CP violation~\cite{Xing:1995jg,Amorim:1998pi,Lipkin:1999qz,Bigi:2012km,Yu:2017oky,Wang:2017gxe} and CPT violation~\cite{Cheng:2021yfr}. The CP violations in the decays $D\rightarrow K_S^0 \pi$ and $\tau\rightarrow\pi K_S^0 \bar{\nu}_{\tau}$ have been reported by Belle~\cite{Ko:2012pe,Belle:2011sna,Ko:2010ng}, BaBar~\cite{delAmoSanchez:2011zza,BABAR:2011aa}, CLEO~\cite{Mendez:2009aa,Dobbs:2007ab} and FOCUS~\cite{Link:2001zj} collaborations, a $2.8\sigma$ discrepancy is observed between the latest BaBar measurement and the Standard Model prediction of the CP asymmetry in the $\tau\rightarrow\pi K_S^0 \bar{\nu}_{\tau}$ decay~\cite{Poireau:2012by,Grossman:2011zk,Cirigliano:2017tqn}. Such a discrepancy has motivated many studies of possible origins of the direct CP asymmetry in $\tau\rightarrow\pi K_S^0 \bar{\nu}_{\tau}$ decay~\cite{Chen:2021udz,Chen:2020uxi,Chen:2019vbr,Dighe:2019odu,Rendon:2019awg,Cirigliano:2019wxv,Castro:2018cot,Delepine:2018amd,Dhargyal:2016kwp,Devi:2013gya,Kimura:2014wsa}. However, the measurements of the CP-violating effect in $\tau\rightarrow\pi K_S^0 \bar{\nu}_{\tau}$ decay still have large uncertainties and no clear conclusion can be drawn in present stage, so more precise data and more decays are needed in both experiment and theory. Moreover, it is crucial to study the CP violation and CPT violation in various reactions, to see the correlations between different processes and probe the sources of CP violation and CPT violation~\cite{Li:2005ci,Karan:2017coa,Karan:2020ada}.

In this paper, we consider the possible CP and CPT asymmetric observations in $K^{\ast}\rightarrow  K_{S,L}^{0}\pi$ decays, in which the possible CP violation and CPT violation are due to $K^0-\bar{K}^0$ oscillation within the Standard Model. Because CP is conserved in the strong decays of $K^{\ast}$ meson into $K\pi$, we can obtain the following properties
\begin{align}
&{\mathcal B}\left(K^{\ast 0}\rightarrow K^0\pi^0\right)={\mathcal B}\left(\bar{K}^{\ast 0}\rightarrow \bar{K}^0\pi^0\right),\label{Eq:kstarkzrocp1}\\
&{\mathcal B}\left(K^{\ast 0}\rightarrow K^+\pi^-\right)={\mathcal B}\left(\bar{K}^{\ast 0}\rightarrow K^-\pi^+\right), \label{Eq:kstarkzrocp2}\\
&{\mathcal B}\left(K^{\ast +}\rightarrow K^0\pi^+\right)={\mathcal B}\left(K^{\ast -}\rightarrow \bar{K}^0\pi^-\right),\label{Eq:kstarkzrocp3}\\
&{\mathcal B}\left(K^{\ast +}\rightarrow K^+\pi^0\right)={\mathcal B}\left(K^{\ast -}\rightarrow K^-\pi^0\right). \label{Eq:kstarkzrocp4}
\end{align}
In the $K^0-\bar{K}^0$ system, the mass eigenstates can be written~\cite{Zyla:2020zbs}
\begin{align}
\left | K^0_L \right\rangle=p\sqrt{1+z}\left | K^0\right\rangle-q\sqrt{1-z}\left | \bar{K}^0\right\rangle,\label{Eq:kldefinition}\\
\left | K^0_S \right\rangle=p\sqrt{1-z}\left | K^0\right\rangle+q\sqrt{1+z}\left | \bar{K}^0\right\rangle,\label{Eq:ksdefinition}
\end{align}
and the corresponding mass eigenbras read~\cite{Capolupo:2011rd,Alvarez-Gaume:1998yzz}
\begin{align}
\left\langle  K^0_L \right|=\frac{q\sqrt{1+z}\left \langle K^0\right| -p\sqrt{1-z}\left \langle \bar{K}^0\right|}{2pq},\label{Eq:kldefinitionbar}\\
\left\langle  K^0_S \right|=\frac{q\sqrt{1-z}\left \langle K^0\right| +p\sqrt{1+z}\left \langle \bar{K}^0\right|}{2pq},\label{Eq:ksdefinitionbar}
\end{align}
where $p$, $q$ and $z$ are complex mixing parameters. If CPT invariance held, we should have $z=0$; If CP and CPT invariance held, we would have $p=q=\sqrt{2}/2$ and $z=0$. The mass and width eigenstates, $K^0_{S,L}$, may also be described with the popular notations
\begin{align}
\left | K^0_L \right\rangle=\frac{1+\epsilon-\delta}{\sqrt{2(1+|\epsilon-\delta|^2)}}\left | K^0\right\rangle-\frac{1-\epsilon+\delta}{\sqrt{2(1+|\epsilon-\delta|^2)}}\left | \bar{K}^0\right\rangle,\label{Eq:kldefinition2}\\
\left | K^0_S \right\rangle=\frac{1+\epsilon+\delta}{\sqrt{2(1+|\epsilon+\delta|^2)}}\left | K^0\right\rangle+\frac{1-\epsilon-\delta}{\sqrt{2(1+|\epsilon+\delta|^2)}}\left | \bar{K}^0\right\rangle,\label{Eq:ksdefinition2}
\end{align}
where the complex parameter $\epsilon$ signifies deviation of the mass eigenstates from the CP eigenstates, $\delta$ is the CPT violating  complex parameter. The parameters $p$, $q$ and $z$ can be expressed in terms of $\epsilon$ and $\delta$ (neglecting terms of $\epsilon\delta$ and $\mathcal{O}(\delta)$)
\begin{align}
p=\frac{1+\epsilon}{\sqrt{2(1+|\epsilon|^2)}},~~~~~~~~~~q=\frac{1-\epsilon}{\sqrt{2(1+|\epsilon|^2)}},~~~~~~~~~~z=-2\delta.\label{Eq:ksdefinition3}
\end{align}
The time-evolved states of the $K^0-\bar{K}^0$ system can be expressed by the mass eigenstates
\begin{align}
\left | K^0_{phys} (t) \right\rangle=\frac{\sqrt{1+z}}{2p} e^{-i m_L t -\frac{1}{2}\Gamma_L t } \left | K^0_L \right\rangle +\frac{\sqrt{1-z}}{2p} e^{-i m_S t -\frac{1}{2}\Gamma_S t } \left | K^0_S \right\rangle,\label{Eq:kzphysdef}\\
\left | \bar{K}^0_{phys} (t) \right\rangle=-\frac{\sqrt{1-z}}{2q} e^{-i m_L t -\frac{1}{2}\Gamma_L t } \left | K^0_L \right\rangle +\frac{\sqrt{1+z}}{2q} e^{-i m_S t -\frac{1}{2}\Gamma_S t } \left | K^0_S \right\rangle.\label{Eq:kbphysdef}
\end{align}
With Eq.(\ref{Eq:kzphysdef}) and Eq.(\ref{Eq:kbphysdef}), the time-dependent amplitudes of the cascade decays $K^{\ast}\rightarrow K^0\pi \rightarrow f_{K^0} \pi$ can be written as
\begin{align}
&A\left(K^{\ast}\rightarrow K^0(t) \pi \rightarrow  f_{K^0} (t) \pi \right)= A\left(K^{\ast}\rightarrow K^0 \pi \right) \cdot A(K^0_{phys} (t)\rightarrow f_{K^0}) ,\label{Eq:amplikstkzptd}
\end{align}
\begin{align}
&A\left(\bar{K}^{\ast}\rightarrow \bar{K}^0(t) \pi \rightarrow  f_{K^0} (t) \pi \right)= A\left(\bar{K}^{\ast}\rightarrow \bar{K}^0 \pi \right) \cdot A(\bar{K}^0_{phys} (t)\rightarrow f_{K^0}) ,\label{Eq:ampliktbkzptd}
\end{align}
where $K^{\ast}$ denotes the $K^{\ast 0}$ (or $K^{\ast +}$) meson, $\bar{K}^{\ast}$ denotes the charge conjugate state of $K^{\ast}$, $f_{K^0}$ denotes the final state from the decay of the $K^0$ or  $\bar{K}^0$ meson. $A(K^0_{phys} (t)\rightarrow f_{K^0})$ and $A(\bar{K}^0_{phys} (t)\rightarrow f_{K^0})$ denotes the amplitude of the $K^0_{phys} (t)\rightarrow f_{K^0}$ and $\bar{K}^0_{phys} (t)\rightarrow f_{K^0}$ decays, respectively, they have the following forms
\begin{align}
&A(K^0_{phys} (t)\rightarrow f_{K^0})=\frac{\sqrt{1+z}}{2p} e^{-i m_L t -\frac{1}{2}\Gamma_L t }  A(K_L^0 \rightarrow f_{K^0} )+\frac{\sqrt{1-z}}{2p} e^{-i m_S t -\frac{1}{2}\Gamma_S t } A(K_S^0 \rightarrow f_{K^0} ),\label{Eq:amtikphystofkt1}\\
&A(\bar{K}^0_{phys} (t)\rightarrow f_{K^0})=-\frac{\sqrt{1-z}}{2q} e^{-i m_L t -\frac{1}{2}\Gamma_L t }  A(K_L^0 \rightarrow f_{K^0} )+\frac{\sqrt{1+z}}{2q} e^{-i m_S t -\frac{1}{2}\Gamma_S t } A(K_S^0 \rightarrow f_{K^0} ).\label{Eq:amtikphystofkt2}
\end{align}
Making use of Eqs.(\ref{Eq:amplikstkzptd}), (\ref{Eq:ampliktbkzptd}), (\ref{Eq:amtikphystofkt1}) and (\ref{Eq:amtikphystofkt2}) and performing integration over phase space, we can obtain
\begin{align}
&{\mathcal B} (K^{\ast}\rightarrow K^0(t) \pi \rightarrow  f_{K^0} (t) \pi )=\frac{{\mathcal B}\left(K^{\ast}\rightarrow K^0\pi\right)}{4 |p|^2}\nonumber\\
&~~~~~~~~~~~~~~~~~~~~~~\cdot\left[\left|\sqrt{1+z}\right|^2 \cdot e^{-\Gamma_L t}\cdot \Gamma(K_L^0\rightarrow f_{K^0})+ \left|\sqrt{1-z}\right|^2 \cdot e^{-\Gamma_S t}\cdot \Gamma(K_S^0\rightarrow f_{K^0})\right.\nonumber\\
&~~~~~~~~~~~~~~~~~~~~~~~~~~\left.+\sqrt{1+z} \cdot \left(\sqrt{1-z}\right)^{*} \cdot e^{-i \Delta m t-\Gamma t}\cdot A^{*}(K_S^0\rightarrow f_{K^0})  \cdot A(K_L^0\rightarrow f_{K^0})\right.\nonumber\\
&~~~~~~~~~~~~~~~~~~~~~~~~~~+\sqrt{1-z} \cdot \left(\sqrt{1+z}\right)^{*} \cdot e^{i \Delta m t-\Gamma t}\cdot A(K_S^0\rightarrow f_{K^0})  \cdot A^{*}(K_L^0\rightarrow f_{K^0})\bigg],
\label{Eq:decaywidthkstt1}
\end{align}
\begin{align}
&{\mathcal B} (\bar{K}^{\ast}\rightarrow \bar{K}^0 (t) \pi \rightarrow  f_{K^0} (t) \pi )=\frac{{\mathcal B}\left(\bar{K}^{\ast}\rightarrow \bar{K}^0 \pi\right)}{4 |q|^2}\nonumber\\
&~~~~~~~~~~~~~~~~~~~~~~\cdot\left[\left|\sqrt{1-z}\right|^2 \cdot e^{-\Gamma_L t}\cdot \Gamma(K_L^0\rightarrow f_{K^0})+ \left|\sqrt{1+z}\right|^2 \cdot e^{-\Gamma_S t}\cdot \Gamma(K_S^0\rightarrow f_{K^0})\right.\nonumber\\
&~~~~~~~~~~~~~~~~~~~~~~~~~~\left.-\sqrt{1-z} \cdot \left(\sqrt{1+z}\right)^{*} \cdot e^{-i \Delta m t-\Gamma t}\cdot A^{*}(K_S^0\rightarrow f_{K^0})  \cdot A(K_L^0\rightarrow f_{K^0})\right.\nonumber\\
&~~~~~~~~~~~~~~~~~~~~~~~~~~-\sqrt{1+z} \cdot \left(\sqrt{1-z}\right)^{*} \cdot e^{i \Delta m t-\Gamma t}\cdot A(K_S^0\rightarrow f_{K^0})  \cdot A^{*}(K_L^0\rightarrow f_{K^0})\bigg],
\label{Eq:decaywidthkstt2}
\end{align}
where $\Delta m$ denotes the difference in masses of $K_L^0$ and $K_S^0$, $\Gamma$ denotes the average in widths of $K_L^0$ and $K_S^0$
\begin{align}
\Delta m=m_L-m_S,~~~~~~~~~~~~~~~~~~~~\Gamma=\frac{\Gamma_L +\Gamma_S}{2}.
\label{Eq:difmasavgam}
\end{align}
The first, second, and the last two terms in the bracket in Eq.(\ref{Eq:decaywidthkstt1}) and Eq.(\ref{Eq:decaywidthkstt2}) are related to the effects of the $K_L^0$ decay, the $K_S^0$ decay and their interference, respectively.

In BESIII, the $K_S^0$ state is reconstructed via its decay into the final states $\pi^+\pi^-$ and a time difference between the $K^{\ast}$ decay and the $K_S^0$ decay in the $K^{\ast}\rightarrow K_S^0 \pi$ decay~\cite{Grossman:2011zk}. By taking into account these experimental features, the branching ratio for the $K^{\ast}\rightarrow K_S^0 \pi$ decay can be defined as
\begin{align}
{\mathcal B}(K^{\ast}\rightarrow K_S^0 \pi)=\frac{\int_{t_0}^{t_1} {\mathcal B}(K^{\ast}\rightarrow K^0(t) \pi \rightarrow   \pi^+\pi^- (t)  \pi ) dt}{\left(e^{-\Gamma_S t_0}-e^{-\Gamma_S t_1}\right)\cdot{\mathcal B}(K^0_S\rightarrow \pi^+\pi^- ) },\label{Eq:kstarkspitimedef}
\end{align}
where $t_0=0.1\tau_S$ and $t_1=2\tau_S\sim 20 \tau_S$ with $\tau_S$ is the $K_S^0$ lifetime, we adopt $t_1=10 \tau_S$ in our calculation. Combining Eq.(\ref{Eq:decaywidthkstt1}) and  Eq.(\ref{Eq:kstarkspitimedef}), we can obtain
\begin{align}
&{\mathcal B}\left(K^{\ast}\rightarrow K^0_S\pi\right)\nonumber\\
&=\frac{{\mathcal B}\left(K^{\ast}\rightarrow K^0\pi\right)}{4\left|p\right|^2 }\cdot\left[\left|\sqrt{1-z}\right|^2 + \left|\sqrt{1+z}\right|^2 \cdot \frac{e^{-\Gamma_L t_0}- e^{-\Gamma_L t_1}}{e^{-\Gamma_S t_0}-e^{-\Gamma_S t_1}} \cdot\frac{ {\mathcal B}(K_L^0\rightarrow \pi^+\pi^-)}{{\mathcal B}(K_S^0\rightarrow \pi^+\pi^-)}\right.\nonumber\\
&+2 Re\left(\sqrt{1+z} \cdot \left(\sqrt{1-z}\right)^{*} \cdot\frac{e^{-i \Delta m t_0-\Gamma t_0}-e^{-i \Delta m t_1-\Gamma t_1}}{e^{-\Gamma_S t_0}-e^{-\Gamma_S t_1}} \cdot  \frac{\Gamma_S}{\Gamma+i \Delta m }\cdot \frac{A(K_L^0\rightarrow \pi^+\pi^-)}{A(K_S^0\rightarrow \pi^+\pi^-)}\right)\bigg],\label{Eq:kstarzksdened1}
\end{align}
Using Eq.(\ref{Eq:kldefinition}), Eq.(\ref{Eq:ksdefinition}) and assuming that the direct CP violation in the $K^0\rightarrow \pi^+\pi^-$ decay can be neglected, we derive
\begin{align}
&\frac{A(K_L^0\rightarrow \pi^+\pi^-)}{A(K_S^0\rightarrow \pi^+\pi^-)}= \frac{p\sqrt{1+z}-q \sqrt{1-z}}{p\sqrt{1-z}+q \sqrt{1+z}}. \label{Eq:ratiokslpipiam}
\end{align}
By combing Eq.(\ref{Eq:ratiokslpipiam}) with Eq.(\ref{Eq:kstarzksdened1}) and introducing the following substitution
\begin{align}
&t_{K^0_S-K_L^0}=\frac{e^{-i \Delta m t_0-\Gamma t_0}-e^{-i \Delta m t_1-\Gamma t_1}}{e^{-\Gamma_S t_0}-e^{-\Gamma_S t_1}}\cdot \frac{\Gamma_S}{\Gamma+i \Delta m },\label{Eq:substitigamma}
\end{align}
we can obtain
\begin{align}
&{\mathcal B}\left(K^{\ast}\rightarrow K^0_S\pi\right)\nonumber\\
&~~=\frac{{\mathcal B}\left(K^{\ast}\rightarrow K^0\pi\right)}{4\left|p\right|^2 }
\cdot\left[\left|\sqrt{1-z}\right|^2 + \left|\sqrt{1+z}\right|^2 \cdot \frac{e^{-\Gamma_L t_0}- e^{-\Gamma_L t_1}}{e^{-\Gamma_S t_0}-e^{-\Gamma_S t_1}} \cdot\frac{ {\mathcal B}(K_L^0\rightarrow \pi^+\pi^-)}{{\mathcal B}(K_S^0\rightarrow \pi^+\pi^-)}\right.\nonumber\\
&~~~~~~~~~~~~~~~~~~~~~~~~~~~~~~~~+2 Re\left(\sqrt{1+z} \cdot \left(\sqrt{1-z}\right)^{*} \cdot t_{K^0_S-K_L^0}\cdot \frac{p\sqrt{1+z}-q \sqrt{1-z}}{p\sqrt{1-z}+q \sqrt{1+z}}\right)\bigg].\label{Eq:kstarzksdened2}
\end{align}
Similarly, we can derive the branching ratio for the $\bar{K}^{\ast}\rightarrow K^0_S\pi$ decay
\begin{align}
&{\mathcal B}\left(\bar{K}^{\ast}\rightarrow K^0_S\pi\right)\nonumber\\
&~~=\frac{{\mathcal B}\left(\bar{K}^{\ast}\rightarrow \bar{K}^0\pi\right)}{4\left|q\right|^2 }\cdot\left[\left|\sqrt{1+z}\right|^2 + \left|\sqrt{1-z}\right|^2 \cdot \frac{e^{-\Gamma_L t_0}- e^{-\Gamma_L t_1}}{e^{-\Gamma_S t_0}-e^{-\Gamma_S t_1}} \cdot\frac{ {\mathcal B}(K_L^0\rightarrow \pi^+\pi^-)}{{\mathcal B}(K_S^0\rightarrow \pi^+\pi^-)}\right.\nonumber\\
&~~~~~~~~~~~~~~~~~~~~~~~~~~~~~~~~-2 Re\left(\sqrt{1-z} \cdot \left(\sqrt{1+z}\right)^{*} \cdot t_{K^0_S-K_L^0}\cdot \frac{p\sqrt{1+z}-q \sqrt{1-z}}{p\sqrt{1-z}+q \sqrt{1+z}}\right)\bigg].\label{Eq:kstbarksdened1}
\end{align}
With the values of the parameters, which are listed in Table~\ref{theparametervalue},
\begin{table}[t]
\begin{center}
\caption{\label{theparametervalue} \small The values of the input parameters used in this paper~\cite{Zyla:2020zbs,Bailey:2018feb}.}
\vspace{0.1cm}
\doublerulesep 0.8pt \tabcolsep 0.18in
\scriptsize
\begin{tabular}{c|c}
\hline
$\text{Re}(\epsilon)=(1.66\pm0.02)\times 10^{-3}$  & $\text{Im}(\epsilon)=(1.57\pm0.02)\times 10^{-3}$   \\
\hline
$|\epsilon|=(2.228\pm0.011)\times 10^{-3}$  & $\Delta m=(3.481\pm0.007)\times 10^{-15} \text{GeV}$   \\
\hline
$\Gamma_L=(1.287\pm0.005)\times 10^{-17} \text{GeV}$ &  $\Gamma_S =(7.351\pm0.003)\times 10^{-15} \text{GeV}$   \\
\hline
${\mathcal B}(K_L^0\rightarrow \pi^+\pi^-)= (1.967\pm0.010)\times 10^{-3}$ & $ {\mathcal B}(K_S^0\rightarrow \pi^+\pi^-)=(69.2\pm0.05)\times 10^{-2} $ \\
\hline
\end{tabular}
\end{center}
\end{table}
we can obtain
\begin{align}
&\frac{e^{-\Gamma_L t_0}- e^{-\Gamma_L t_1}}{e^{-\Gamma_S t_0}-e^{-\Gamma_S t_1}} \cdot\frac{ {\mathcal B}(K_L^0\rightarrow \pi^+\pi^-)}{{\mathcal B}(K_S^0\rightarrow \pi^+\pi^-)}= (5.40\pm 0.03)\times 10^{-5}. \label{Eq:valtimeksklbrratio}
\end{align}
Combing Eq.(\ref{Eq:kstarzksdened2}), Eq.(\ref{Eq:kstbarksdened1}) and Eq.(\ref{Eq:valtimeksklbrratio}), we can obtain
\begin{align}
&{\mathcal B}\left(K^{\ast}\rightarrow K^0_S\pi\right)=\frac{{\mathcal B}\left(K^{\ast}\rightarrow K^0\pi\right)}{4\left|p\right|^2 }\nonumber\\
&~~~~~~~~~~~~~~~\cdot\left[\left|\sqrt{1-z}\right|^2 +2 Re\left(\sqrt{1+z} \cdot \left(\sqrt{1-z}\right)^{*} \cdot t_{K^0_S-K_L^0}\cdot \frac{p\sqrt{1+z}-q \sqrt{1-z}}{p\sqrt{1-z}+q \sqrt{1+z}}\right)\right],\label{Eq:kstarzksdecay}
\end{align}
\begin{align}
&{\mathcal B}\left(\bar{K}^{\ast}\rightarrow K^0_S\pi\right)=\frac{{\mathcal B}\left(\bar{K}^{\ast}\rightarrow \bar{K}^0\pi\right)}{4\left|q\right|^2 }\nonumber\\
&~~~~~~~~~~~~~~~\cdot\left[\left|\sqrt{1+z}\right|^2 -2 Re\left(\sqrt{1-z} \cdot \left(\sqrt{1+z}\right)^{*} \cdot t_{K^0_S-K_L^0}\cdot \frac{p\sqrt{1+z}-q \sqrt{1-z}}{p\sqrt{1-z}+q \sqrt{1+z}}\right)\right].\label{Eq:kstarzbarksdecay}
\end{align}
In BESIII, the $K_L^0$ state is defined via a large time difference between the $K^{\ast}$ decay and the $K_L^0$ decay and mostly decay outside the detector in the $K^{\ast}\rightarrow K^0_L\pi$ decay~\cite{BESIII:2020nme,BESIII:2021yam,Asner:2008nq}. Basing on these experimental features, the branching ratio for the $K^{\ast}\rightarrow K^0_L\pi$ decay can be defined as
\begin{align}
&{\mathcal B}\left(K^{\ast}\rightarrow K^0_L\pi\right)=\frac{\int_{t_2}^{+\infty}  {\mathcal B}(K^{\ast}\rightarrow K^0(t) \pi \rightarrow   f_{K_L^0} (t)  \pi ) dt}{e^{-\Gamma_L t_2}\cdot{\mathcal B}(K^0_L\rightarrow f_{K_L^0} ) },\label{Eq:kstarklpibrdef1}
\end{align}
where $f_{K_L^0}$ denotes the final state of the $K_L^0$ decay, $t_2\geq 100\tau_S$. Combining Eq.(\ref{Eq:decaywidthkstt1}) and  Eq.(\ref{Eq:kstarklpibrdef1}), we can obtain
\begin{align}
&{\mathcal B}\left(K^{\ast}\rightarrow K^0_L\pi\right)\nonumber\\
&~~~~=\frac{{\mathcal B}\left(K^{\ast}\rightarrow K^0\pi\right)}{4\left|p\right|^2 }\cdot\left[\left|\sqrt{1+z}\right|^2 + \left|\sqrt{1-z}\right|^2 \cdot e^{-(\Gamma_S-\Gamma_L) t_2}\cdot\frac{ {\mathcal B}(K_S^0\rightarrow f_{K_L^0})}{{\mathcal B}(K_L^0\rightarrow f_{K_L^0})}\right.\nonumber\\
&~~~~~~+2 Re\left(\sqrt{1-z} \cdot \left(\sqrt{1+z}\right)^{*} \cdot\frac{e^{i \Delta m t_2-\frac{\Gamma_S-\Gamma_L}{2} t_2}}{\Gamma-i \Delta m} \cdot \frac{A(K_S^0\rightarrow f_{K_L^0})}{A(K_L^0\rightarrow f_{K_L^0})}\right)\bigg].\label{Eq:kstarklpibrdef2}
\end{align}
Using the values of the parameters in Table~\ref{theparametervalue}, we can obtain
\begin{align}
&e^{-(\Gamma_S-\Gamma_L) t_2}\le 4.4\times 10^{-44},
~~~~~~~~~~~~~~~~~~~~~~~~~~e^{-\frac{\Gamma_S-\Gamma_L}{2} t_2}\le 2.1\times 10^{-22},\label{Eq:valuettexpimp}
\end{align}
so the second and the third terms in the bracket in Eq.(\ref{Eq:kstarklpibrdef2}), which correspond respectively to the effects of the $K_S^0$ decay and the interference between the $K_S^0$ decay and the $K_L^0$ decay, can be neglected, then we obtain
\begin{align}
&{\mathcal B}\left(K^{\ast}\rightarrow K^0_L\pi\right)={\mathcal B}\left(K^{\ast}\rightarrow K^0\pi\right)\frac{\left|\sqrt{1+z}\right|^2}{4\left|p\right|^2 }.\label{Eq:kstarzkldecay}
\end{align}
Similarly, we can derive the branching ratio for the $\bar{K}^{\ast}\rightarrow K^0_L\pi$ decay
\begin{align}
&{\mathcal B}\left(\bar{K}^{\ast}\rightarrow K^0_L\pi\right)={\mathcal B}\left(\bar{K}^{\ast}\rightarrow \bar{K}^0\pi\right) \frac{\left|\sqrt{1-z}\right|^2}{4\left|q\right|^2 },\label{Eq:kstarzbarkldecay}
\end{align}

By combining Eqs.(\ref{Eq:kstarkzrocp1})-(\ref{Eq:kstarkzrocp4}), Eqs.(\ref{Eq:kstarzksdecay})-(\ref{Eq:kstarzbarksdecay}) and Eqs.(\ref{Eq:kstarzkldecay})-(\ref{Eq:kstarzbarkldecay}), and assuming $z=0$, we can derive the following observations of CP asymmetry
\begin{align}
&{\mathcal A}_{CP}\left(K^{\ast}\rightarrow K^0_{S}\pi\right)=\frac{{\mathcal B}\left(K^{\ast}\rightarrow K^0_{S}\pi\right)-{\mathcal B}\left(\bar{K}^{\ast}\rightarrow K^0_{S}\pi\right)}{{\mathcal B}\left(K^{\ast}\rightarrow K^0_{S}\pi\right)+{\mathcal B}\left(\bar{K}^{\ast}\rightarrow K^0_{S}\pi\right)}
=\frac{\left|q\right|^2-\left|p\right|^2}{\left|q\right|^2+\left|p\right|^2}+2 Re\left( t_{K^0_S-K_L^0}\cdot \frac{p-q}{p+q}\right),\label{Eq:kstarkscpasy}
\end{align}
and
\begin{align}
&{\mathcal A}_{CP}\left(K^{\ast}\rightarrow K^0_{L}\pi\right)=\frac{{\mathcal B}\left(K^{\ast}\rightarrow K^0_{L}\pi\right)-{\mathcal B}\left(\bar{K}^{\ast}\rightarrow K^0_{L}\pi\right)}{{\mathcal B}\left(K^{\ast}\rightarrow K^0_{L}\pi\right)+{\mathcal B}\left(\bar{K}^{\ast}\rightarrow K^0_{L}\pi\right)}=\frac{\left|q\right|^2-\left|p\right|^2}{\left|q\right|^2+\left|p\right|^2},\label{Eq:kstarklcpasy}
\end{align}
where $K^{\ast}$ denotes the $K^{\ast 0}$ (or $K^{\ast +}$) meson, $\bar{K}^{\ast}$ denotes the charge conjugate state of $K^{\ast}$.

The $K^{\ast}$ meson can be produced through the $J/\psi$ decays at BESIII. The CP asymmetry observables can be defined in the $J/\psi$ decays which involve $K^{\ast}$ meson in the final states
\begin{align}
&{\mathcal A}_{CP}\left(J/\psi\rightarrow K^{\ast}f_{J/\psi}\rightarrow K^0_{S,L}\pi f_{J/\psi}\right)\nonumber\\
&~=\frac{{\mathcal B}\left(J/\psi\rightarrow K^{\ast} f_{J/\psi}\rightarrow K^0_{S,L} \pi f_{J/\psi}\right)-{\mathcal B}\left(J/\psi\rightarrow \bar{K}^{\ast} \bar{f}_{J/\psi}\rightarrow K^0_{S,L} \pi \bar{f}_{J/\psi}\right)}{{\mathcal B}\left(J/\psi\rightarrow K^{\ast} f_{J/\psi}\rightarrow K^0_{S,L}\pi f_{J/\psi}\right)+{\mathcal B}\left(J/\psi\rightarrow \bar{K}^{\ast} \bar{f}_{J/\psi}\rightarrow K^0_{S,L} \pi \bar{f}_{J/\psi}\right)},\label{Eq:jpsikstarkslcp}
\end{align}
where $f_{J/\psi}$ denotes the final state ($K^{\ast}$ excepted) in the $J/\psi$ decays, $\bar{f}_{J/\psi}$ is the charge conjugate state of $f_{J/\psi}$. For example, the CP asymmetry observables can be defined in $J/\psi\rightarrow \gamma K^{\ast 0} \bar{K}^{\ast 0}$ and $J/\psi\rightarrow K^\pm K^{\ast \mp}$ decays
\begin{align}
&{\mathcal A}_{CP}\left(J/\psi\rightarrow \gamma K^{\ast 0} \bar{K}^{\ast 0}\rightarrow \gamma K^0_{S,L}\pi^0 K^-\pi^+ \right)\nonumber\\
&=\frac{{\mathcal B}\left(J/\psi\rightarrow \gamma K^{\ast 0} \bar{K}^{\ast 0}\rightarrow \gamma K^0_{S,L}\pi^0 K^-\pi^+\right)-{\mathcal B}\left(J/\psi\rightarrow \gamma K^{\ast 0} \bar{K}^{\ast 0}\rightarrow \gamma K^+\pi^- K^0_{S,L}\pi^0 \right)}{{\mathcal B}\left(J/\psi\rightarrow \gamma K^{\ast 0} \bar{K}^{\ast 0}\rightarrow \gamma K^0_{S,L}\pi^0 K^-\pi^+\right)+{\mathcal B}\left(J/\psi\rightarrow \gamma K^{\ast 0} \bar{K}^{\ast 0}\rightarrow \gamma K^+\pi^- K^0_{S,L}\pi^0 \right)},\label{Eq:jpsigakstarkstarcp}
\end{align}
\begin{align}
&{\mathcal A}_{CP}\left(J/\psi\rightarrow K^\pm K^{\ast \mp}\rightarrow K^\pm K^0_{S,L}\pi^\mp\right)\nonumber\\
&~~~~~~~~~~~~=\frac{{\mathcal B}\left(J/\psi\rightarrow K^- K^{\ast +}\rightarrow K^- K^0_{S,L}\pi^+\right)-{\mathcal B}\left(J/\psi\rightarrow K^+ K^{\ast -}\rightarrow K^+ K^0_{S,L}\pi^-\right)}{{\mathcal B}\left(J/\psi\rightarrow K^- K^{\ast +}\rightarrow K^- K^0_{S,L}\pi^+\right)+{\mathcal B}\left(J/\psi\rightarrow K^+ K^{\ast -}\rightarrow K^+ K^0_{S,L}\pi^-\right)}.\label{Eq:jpsigakpmkstarcp}
\end{align}
According to Eq.(\ref{Eq:kstarkzrocp2}), Eq.(\ref{Eq:kstarkzrocp4}) and Eqs.(\ref{Eq:kstarkscpasy})-(\ref{Eq:jpsikstarkslcp}), we can derive
\begin{align}
{\mathcal A}_{CP}\left(J/\psi\rightarrow K^{\ast}f_{J/\psi}\rightarrow K^0_{S,L}\pi f_{J/\psi}\right)={\mathcal A}_{CP}\left(K^{\ast}\rightarrow K^0_{S,L}\pi\right).\label{Eq:jpsieqkstarcp}
\end{align}
By using Eq.(\ref{Eq:jpsikstarkslcp}) and Eq.(\ref{Eq:jpsieqkstarcp}), the number of the observed signal events on the CP violation in $J/\psi$ decays can be derived
\begin{align}
&N_{CP}^{K^0_{S,L}}=\left|{\mathcal A}_{CP}\left(K^{\ast}\rightarrow K^0_{S,L}\pi\right)\right| \cdot N_{J/\psi}\cdot \varepsilon_{K^0_{S,L}} \nonumber\\
&~~~~~~~~~~~\cdot {\mathcal B}\left(J/\psi\rightarrow K^{\ast} f_{J/\psi}+c.c.\rightarrow K^0_{S,L}\pi f_{J/\psi}+c.c.\rightarrow f_{K^0_{S,L}} \pi f_{J/\psi}+c.c.\right) \nonumber\\
&~~~\approx 2\left|{\mathcal A}_{CP}\left(K^{\ast}\rightarrow K^0_{S,L}\pi\right)\right| \cdot N_{J/\psi}\cdot \varepsilon_{K^0_{S,L}} \cdot {\mathcal B}\left(J/\psi\rightarrow K^{\ast} f_{J/\psi}\rightarrow K^0_{S,L}\pi f_{J/\psi}\rightarrow f_{K^0_{S,L}} \pi f_{J/\psi}\right),\label{Eq:numbercpksl}
\end{align}
where $N_{J/\psi}$ is the number of $J/\psi$ events accumulated at BESIII. During several run periods from 2009 to 2019, a total data sample of $10^{10}$ $J/\psi$ events was collected with BESIII detector~\cite{BESIII:2012pbg,BESIII:2016kpv,BESIII:2020nme}. Moreover, accelerators at the tau-charm energy region with luminosity 100 times higher than BEPCII are being proposed~\cite{Yuan:2021yks,STCFzhao,SCTFlevichev}, the detectors in these new facilities will be able to collect $10^{12}$ $J/\psi$ events in one year's running time. In this paper, we will perform the calculation under two cases: $N_{J/\psi}=10^{10}$ and $N_{J/\psi}=10^{12}$. $\varepsilon_{K^0_{S,L}}$ in Eq.(\ref{Eq:numbercpksl}) are the selection efficiencies of $J/\psi$ decays at BESIII. $f_{K^0_{S,L}}$ in Eq.(\ref{Eq:numbercpksl}) denote the final states in the $K^0_{S,L}$ decays. Here, it is worth noting that the $K^0_{L}$ meson couldn't be reconstructed by its decays because of its large life\cite{BESIII:2020nme,BESIII:2021yam,Asner:2008nq}. Instead, all particles except for the $K^0_{L}$ are reconstructed and the presence of a $K^0_{L}$ can be inferred from the missing four momentum in the decays contain a $K^0_{L}$ in the final states. So we don't consider the branching ratios of $K^0_{L}$ decays in the calculation.

The branching ratios of the $J/\psi\rightarrow K^{\ast} f_{J/\psi}\rightarrow K^0_{S,L}\pi f_{J/\psi}$ decays can be expressed as the products of the branching ratios of the $J/\psi\rightarrow K^{\ast} f_{J/\psi}$ decays and the $K^{\ast} \rightarrow K^0_{S,L}\pi$ decays. The branching ratios of the $J/\psi\rightarrow K^{\ast} f_{J/\psi}$ decays can be taken directly from the Particle Data Group (PDG)~\cite{Zyla:2020zbs} and Ref.~\cite{BaBar:2017pkz}. For the branching ratios of the $K^{\ast} \rightarrow K^0_{S,L}\pi$ decays, we apply the PDG result ${\mathcal B}\left( K^{\ast}\rightarrow K \pi \right)\approx100\%$ and isospin relation to obtain
\begin{align}
&{\mathcal B}\left(K^{\ast 0}\rightarrow K^0\pi^0\right)=\frac{1}{3},~~~~~~~~{\mathcal B}\left(K^{\ast 0}\rightarrow K^+\pi^-\right)=\frac{2}{3},\label{Eq:kstarkzrocp5}\\
&{\mathcal B}\left(\bar{K}^{\ast 0}\rightarrow \bar{K}^0\pi^0\right)=\frac{1}{3},~~~~~~~~{\mathcal B}\left(\bar{K}^{\ast 0}\rightarrow K^-\pi^+\right)=\frac{2}{3}, \label{Eq:kstarkzrocp6}\\
&{\mathcal B}\left(K^{\ast +}\rightarrow K^0\pi^+\right)=\frac{2}{3},~~~~~~~~{\mathcal B}\left(K^{\ast +}\rightarrow K^+\pi^0\right)=\frac{1}{3},\label{Eq:kstarkzrocp7}\\
&{\mathcal B}\left(K^{\ast -}\rightarrow \bar{K}^0\pi^-\right)=\frac{2}{3},~~~~~~~~{\mathcal B}\left(K^{\ast -}\rightarrow K^-\pi^0\right)=\frac{1}{3}. \label{Eq:kstarkzrocp8}
\end{align}
Then, from Eqs.(\ref{Eq:kstarkzrocp5})-(\ref{Eq:kstarkzrocp8}) and under the assumptions of CP conservation and CPT conservation, we have
\begin{align}
&{\mathcal B}\left(K^{\ast 0}\rightarrow K^0_S\pi^0\right)=\frac{1}{6},~~~~~~~~{\mathcal B}\left(K^{\ast 0}\rightarrow K^0_L\pi^0\right)=\frac{1}{6},\label{Eq:kstarkslcp1}\\
&{\mathcal B}\left(\bar{K}^{\ast 0}\rightarrow K^0_S\pi^0\right)=\frac{1}{6},~~~~~~~~{\mathcal B}\left(\bar{K}^{\ast 0}\rightarrow K^0_L\pi^0\right)=\frac{1}{6}, \label{Eq:kstarkslcp2}\\
&{\mathcal B}\left(K^{\ast +}\rightarrow K^0_S\pi^+\right)=\frac{1}{3},~~~~~~~~{\mathcal B}\left(K^{\ast +}\rightarrow K^0_L\pi^+\right)=\frac{1}{3},\label{Eq:kstarkslcp3}\\
&{\mathcal B}\left(K^{\ast -}\rightarrow K^0_S\pi^-\right)=\frac{1}{3},~~~~~~~~{\mathcal B}\left(K^{\ast -}\rightarrow K^0_L\pi^-\right)=\frac{1}{3}. \label{Eq:kstarkslcp4}
\end{align}
According to Eqs.(\ref{Eq:kstarkzrocp5})-(\ref{Eq:kstarkslcp4}) and the values for the branching ratios of the $J/\psi\rightarrow K^{\ast} f_{J/\psi}$ and $K^0_{S}\rightarrow f_{K^0_{S}}$ decays which are taken directly from PDG~\cite{Zyla:2020zbs} and Ref.~\cite{BaBar:2017pkz}, we collect the branching ratios of $J/\psi\rightarrow K^{\ast} f_{J/\psi}\rightarrow K^0_{S,L}\pi f_{J/\psi}\rightarrow f_{K^0_{S,L}}\pi f_{J/\psi}$ decays, which are shown in Table~\ref{totbranchrakstar}.
\begin{table}[t]
\begin{center}
\caption{\label{totbranchrakstar} \small The branching fractions of $J/\psi\rightarrow K^{\ast} f_{J/\psi}\rightarrow K^0_{S,L}\pi f_{J/\psi}\rightarrow f_{K^0_{S,L}}\pi f_{J/\psi}$ decays.}
\vspace{0.1cm}
\doublerulesep 0.8pt \tabcolsep 0.18in
\scriptsize
\begin{tabular}{c|c}
\hline
the branching ratio for the decay channel & the numerical result  \\
\hline
${\mathcal B}\left( J/\psi\rightarrow  K^{\ast 0} \bar{K}^{\ast 0}\rightarrow K^0_{S}\pi^0 K^-\pi^+ \rightarrow \pi^+\pi^-\pi^0 K^-\pi^+\right)$ & $\left(1.77\pm{0.46}\right)\times 10^{-5}$   \\
\hline
${\mathcal B}\left( J/\psi\rightarrow  K^{\ast 0} \bar{K}^{\ast 0}\rightarrow K^0_{L}\pi^0 K^-\pi^+\right)$ & $\left(2.56\pm{0.67}\right)\times 10^{-5}$   \\
\hline
${\mathcal B}\left( J/\psi\rightarrow  K^{\ast +} K^{\ast -}\rightarrow K^0_{S}\pi^+ K^-\pi^0 \rightarrow \pi^+\pi^-\pi^+ K^-\pi^0\right)$ & $\left(7.69^{-3.08}_{+1.69}\right)\times 10^{-5}$   \\
\hline
${\mathcal B}\left( J/\psi\rightarrow  K^{\ast +} K^{\ast -}\rightarrow K^0_{L}\pi^+ K^-\pi^0 \right)$ & $\left(1.11^{-0.44}_{+0.24}\right)\times 10^{-4}$    \\
\hline
${\mathcal B}\left( J/\psi\rightarrow \eta K^{\ast 0} \bar{K}^{\ast 0}\rightarrow \eta K^0_{S}\pi^0 K^-\pi^+ \rightarrow \gamma \gamma \pi^+\pi^-\pi^0 K^-\pi^+\right)$ & $\left(3.48\pm{0.79}\right)\times 10^{-5}$   \\
\hline
${\mathcal B}\left( J/\psi\rightarrow \eta K^{\ast 0} \bar{K}^{\ast 0}\rightarrow \gamma \gamma K^0_{L}\pi^0 K^-\pi^+ \right)$ & $\left(5.04\pm{1.14}\right)\times 10^{-5}$   \\
\hline
${\mathcal B}\left( J/\psi\rightarrow \gamma K^{\ast 0} \bar{K}^{\ast 0}\rightarrow \gamma K^0_{S}\pi^0 K^-\pi^+ \rightarrow \gamma \pi^+\pi^-\pi^0 K^-\pi^+\right)$ & $\left(3.08\pm{1.00}\right)\times 10^{-4}$   \\
\hline
${\mathcal B}\left( J/\psi\rightarrow \gamma K^{\ast 0} \bar{K}^{\ast 0}\rightarrow \gamma K^0_{L}\pi^0 K^-\pi^+ \right)$ & $\left(4.44\pm{1.44}\right)\times 10^{-4}$   \\
\hline
${\mathcal B}\left( J/\psi\rightarrow  \eta^{\prime} K^{\ast +}K^- \rightarrow \eta^{\prime} K^0_{S}\pi^+ K^- \rightarrow \pi^+\pi^-\eta K^0_{S}\pi^+ K^-\rightarrow\pi^+\pi^-\gamma\gamma \pi^+ \pi^- \pi^+ K^-\right)$ & $\left(2.86\pm{0.25}\right)\times 10^{-5}$   \\
\hline
${\mathcal B}\left( J/\psi\rightarrow  \eta^{\prime} K^{\ast +}K^- \rightarrow \pi^+\pi^-\eta K^0_{L}\pi^+ K^- \rightarrow \pi^+\pi^-\gamma\gamma K^0_{L}\pi^+ K^-\right)$ & $\left(4.13\pm{0.37}\right)\times 10^{-5}$   \\
\hline
${\mathcal B}\left( J/\psi\rightarrow  K^{\ast +}K^- \rightarrow  K^0_{S}\pi^+ K^- \rightarrow \pi^+\pi^-\pi^+ K^-\right)$ & $\left(6.92^{-1.15}_{+0.92}\right)\times 10^{-4}$   \\
\hline
${\mathcal B}\left( J/\psi\rightarrow  K^{\ast +} K^- \rightarrow  K^0_{L}\pi^+ K^- \right)$ & $\left(1.00^{-0.17}_{+0.13}\right)\times 10^{-3}$   \\
\hline
${\mathcal B}\left( J/\psi\rightarrow  K^{\ast 0} K^-\pi^+\rightarrow K^0_{S}\pi^0 K^-\pi^+ \rightarrow \pi^+\pi^-\pi^0 K^-\pi^+\right)$ & $\left(4.44\pm{0.92}\right)\times 10^{-4}$   \\
\hline
${\mathcal B}\left( J/\psi\rightarrow  K^{\ast 0} K^-\pi^+\rightarrow K^0_{L}\pi^0 K^-\pi^+ \right)$ & $\left(6.42\pm{1.33}\right)\times 10^{-4}$   \\
\hline
${\mathcal B}\left( J/\psi\rightarrow  K^{\ast +} K^-\pi^0\rightarrow K^0_{S}\pi^+ K^-\pi^0 \rightarrow \pi^+\pi^-\pi^+ K^-\pi^0\right)$ & $\left(3.46\pm{1.11}\right)\times 10^{-4}$   \\
\hline
${\mathcal B}\left( J/\psi\rightarrow  K^{\ast +} K^-\pi^0\rightarrow K^0_{L}\pi^+ K^-\pi^0 \right)$ & $\left(5.00\pm{1.57}\right)\times 10^{-4}$   \\
\hline
${\mathcal B}\left( J/\psi\rightarrow  \omega K^{\ast +}K^- \rightarrow \omega K^0_{S}\pi^+ K^- \rightarrow \pi^+\pi^-\pi^0 \pi^+\pi^-\pi^+ K^-\right)$ & $\left(6.28\pm0.93\right)\times 10^{-4}$   \\
\hline
${\mathcal B}\left( J/\psi\rightarrow  \omega K^{\ast +} K^- \rightarrow  \pi^+\pi^-\pi^0 K^0_{L}\pi^+ K^- \right)$ & $\left(9.07\pm1.34\right)\times 10^{-4}$   \\
\hline
\end{tabular}
\end{center}
\end{table}
Here, we only consider the decay channels with a branching ratios larger than $1.0\times 10^{-5}$. With the values of the parameters in Table~\ref{theparametervalue} and combining Eqs.(\ref{Eq:ksdefinition3}), (\ref{Eq:substitigamma}), (\ref{Eq:kstarkscpasy}) and (\ref{Eq:kstarklcpasy}), we can obtain
\begin{align}
&{\mathcal A}_{CP}\left(K^{\ast}\rightarrow K^0_{S}\pi\right)=\left(3.64\pm0.04\right)\times 10^{-3},\label{Eq:acpkstarkspi}\\
&{\mathcal A}_{CP}\left(K^{\ast}\rightarrow K^0_{L}\pi\right)=\left(-3.32\pm0.04\right)\times 10^{-3}. \label{Eq:acpkstarklpi}
\end{align}
According to the Table~\ref{totbranchrakstar} and combining Eqs.(\ref{Eq:numbercpksl}), (\ref{Eq:acpkstarkspi}) and (\ref{Eq:acpkstarklpi}), we derive the numerical results of $N_{CP}^{K^0_{S,L}}$, which are shown in Table~\ref{expectnumevent}, based on the samples of $10^{10}$ and $10^{12}$ $J/\psi$ events. Obviously, the CP violations in $J/\psi$ decays with $K^{\ast}$ meson in the final states can be unambiguously observed at BESIII.
\begin{table}[t]
\newcommand{\tabincell}[2]{\begin{tabular}{@{}#1@{}}#2\end{tabular}}
\begin{center}
\caption{\label{expectnumevent} \small The expected numbers of the observed signal events on the CP violation in $J/\psi$ decays.}
\vspace{0.1cm}
\doublerulesep 0.8pt \tabcolsep 0.18in
\scriptsize
\begin{tabular}{c|c}
\hline
the decay channel & $N_{CP}^{K^0_{S,L}}$ \\
\hline
$ J/\psi\rightarrow  K^{\ast 0} \bar{K}^{\ast 0}\rightarrow K^0_{S}\pi^0 K^-\pi^+ \rightarrow \pi^+\pi^-\pi^0 K^-\pi^+$ & \tabincell{c}{$N_{J/\psi}=10^{10}$: $(1.3\pm 0.3)\times 10^{3} \times\varepsilon_{K^0_{S}} $\\$N_{J/\psi}=10^{12}$: $(1.3\pm 0.3)\times 10^{5} \times\varepsilon_{K^0_{S}} $}  \\
\hline
$ J/\psi\rightarrow  K^{\ast 0} \bar{K}^{\ast 0}\rightarrow K^0_{L}\pi^0 K^-\pi^+$ & \tabincell{c}{$N_{J/\psi}=10^{10}$: $(1.7\pm0.4)\times 10^{3} \times\varepsilon_{K^0_{L}} $\\$N_{J/\psi}=10^{12}$: $(1.7\pm0.4)\times 10^{5} \times\varepsilon_{K^0_{L}} $}   \\
\hline
$ J/\psi\rightarrow  K^{\ast +} K^{\ast -}\rightarrow K^0_{S}\pi^+ K^-\pi^0 \rightarrow \pi^+\pi^-\pi^+ K^-\pi^0$ & \tabincell{c}{$N_{J/\psi}=10^{10}$: $(5.6^{-2.2}_{+1.2})\times 10^{3} \times\varepsilon_{K^0_{S}} $\\$N_{J/\psi}=10^{12}$: $(5.6^{-2.2}_{+1.2})\times 10^{5} \times\varepsilon_{K^0_{S}} $}   \\
\hline
$J/\psi\rightarrow  K^{\ast +} K^{\ast -}\rightarrow K^0_{L}\pi^+ K^-\pi^0 $ & \tabincell{c}{$N_{J/\psi}=10^{10}$: $(7.4^{-3.0}_{+1.6})\times 10^{3} \times\varepsilon_{K^0_{L}} $\\$N_{J/\psi}=10^{12}$: $(7.4^{-3.0}_{+1.6})\times 10^{5} \times\varepsilon_{K^0_{L}} $}   \\
\hline
$ J/\psi\rightarrow \eta K^{\ast 0} \bar{K}^{\ast 0}\rightarrow \eta K^0_{S}\pi^0 K^-\pi^+ \rightarrow \gamma \gamma \pi^+\pi^-\pi^0 K^-\pi^+$ & \tabincell{c}{$N_{J/\psi}=10^{10}$: $(2.5\pm0.6)\times 10^{3} \times\varepsilon_{K^0_{S}} $\\$N_{J/\psi}=10^{12}$: $(2.5\pm0.6)\times 10^{5} \times\varepsilon_{K^0_{S}} $}   \\
\hline
$ J/\psi\rightarrow \eta K^{\ast 0} \bar{K}^{\ast 0}\rightarrow \gamma \gamma K^0_{L}\pi^0 K^-\pi^+ $ & \tabincell{c}{$N_{J/\psi}=10^{10}$: $(3.3\pm0.8)\times 10^{3} \times\varepsilon_{K^0_{L}} $\\$N_{J/\psi}=10^{12}$: $(3.3\pm0.8)\times 10^{5} \times\varepsilon_{K^0_{L}} $}   \\
\hline
$J/\psi\rightarrow \gamma K^{\ast 0} \bar{K}^{\ast 0}\rightarrow \gamma K^0_{S}\pi^0 K^-\pi^+ \rightarrow \gamma \pi^+\pi^-\pi^0 K^-\pi^+$ & \tabincell{c}{$N_{J/\psi}=10^{10}$: $(2.2\pm0.7)\times 10^{4} \times\varepsilon_{K^0_{S}} $\\$N_{J/\psi}=10^{12}$: $(2.2\pm0.7)\times 10^{6} \times\varepsilon_{K^0_{S}} $}   \\
\hline
$ J/\psi\rightarrow \gamma K^{\ast 0} \bar{K}^{\ast 0}\rightarrow \gamma K^0_{L}\pi^0 K^-\pi^+ $ & \tabincell{c}{$N_{J/\psi}=10^{10}$: $(3.0\pm1.0)\times 10^{4} \times\varepsilon_{K^0_{L}} $\\$N_{J/\psi}=10^{12}$: $(3.0\pm1.0)\times 10^{6} \times\varepsilon_{K^0_{L}} $}   \\
\hline
$ J/\psi\rightarrow  \eta^{\prime} K^{\ast +}K^-  \rightarrow \pi^+\pi^-\eta K^0_{S}\pi^+ K^-\rightarrow\pi^+\pi^-\gamma\gamma \pi^+ \pi^- \pi^+ K^-$ & \tabincell{c}{$N_{J/\psi}=10^{10}$: $(2.1\pm0.2)\times 10^{3} \times\varepsilon_{K^0_{S}} $\\$N_{J/\psi}=10^{12}$: $(2.1\pm0.2)\times 10^{5} \times\varepsilon_{K^0_{S}} $}   \\
\hline
$ J/\psi\rightarrow  \eta^{\prime} K^{\ast +}K^- \rightarrow \pi^+\pi^-\eta K^0_{L}\pi^+ K^- \rightarrow \pi^+\pi^-\gamma\gamma K^0_{L}\pi^+ K^-$ & \tabincell{c}{$N_{J/\psi}=10^{10}$: $(2.7\pm0.2)\times 10^{3} \times\varepsilon_{K^0_{L}} $\\$N_{J/\psi}=10^{12}$: $(2.7\pm0.2)\times 10^{5} \times\varepsilon_{K^0_{L}} $}   \\
\hline
$J/\psi\rightarrow  K^{\ast +}K^- \rightarrow  K^0_{S}\pi^+ K^- \rightarrow \pi^+\pi^-\pi^+ K^-$ & \tabincell{c}{$N_{J/\psi}=10^{10}$: $(5.0^{-0.8}_{+0.7})\times 10^{4} \times\varepsilon_{K^0_{S}} $\\$N_{J/\psi}=10^{12}$: $(5.0^{-0.8}_{+0.7})\times 10^{6} \times\varepsilon_{K^0_{S}} $}   \\
\hline
$J/\psi\rightarrow  K^{\ast +} K^- \rightarrow  K^0_{L}\pi^+ K^- $ & \tabincell{c}{$N_{J/\psi}=10^{10}$: $(6.6^{-1.1}_{+0.9})\times 10^{4} \times\varepsilon_{K^0_{L}} $\\$N_{J/\psi}=10^{12}$: $(6.6^{-1.1}_{+0.9})\times 10^{6} \times\varepsilon_{K^0_{L}} $}   \\
\hline
$J/\psi\rightarrow  K^{\ast 0} K^-\pi^+\rightarrow K^0_{S}\pi^0 K^-\pi^+ \rightarrow \pi^+\pi^-\pi^0 K^-\pi^+$ & \tabincell{c}{$N_{J/\psi}=10^{10}$: $(3.2\pm0.7)\times 10^{4} \times\varepsilon_{K^0_{S}} $\\$N_{J/\psi}=10^{12}$: $(3.2\pm0.7)\times 10^{6} \times\varepsilon_{K^0_{S}} $}   \\
\hline
$ J/\psi\rightarrow  K^{\ast 0} K^-\pi^+\rightarrow K^0_{L}\pi^0 K^-\pi^+ $ & \tabincell{c}{$N_{J/\psi}=10^{10}$: $(4.3\pm0.9)\times 10^{4} \times\varepsilon_{K^0_{L}} $\\$N_{J/\psi}=10^{12}$: $(4.3\pm0.9)\times 10^{6} \times\varepsilon_{K^0_{L}} $}   \\
\hline
$ J/\psi\rightarrow  K^{\ast +} K^-\pi^0\rightarrow K^0_{S}\pi^+ K^-\pi^0 \rightarrow \pi^+\pi^-\pi^+ K^-\pi^0$ & \tabincell{c}{$N_{J/\psi}=10^{10}$: $(2.5\pm0.8)\times 10^{4} \times\varepsilon_{K^0_{S}} $\\$N_{J/\psi}=10^{12}$: $(2.5\pm0.8)\times 10^{6} \times\varepsilon_{K^0_{S}} $}   \\
\hline
$ J/\psi\rightarrow  K^{\ast +} K^-\pi^0\rightarrow K^0_{L}\pi^+ K^-\pi^0 $ & \tabincell{c}{$N_{J/\psi}=10^{10}$: $(3.3\pm1.0)\times 10^{4} \times\varepsilon_{K^0_{L}} $\\$N_{J/\psi}=10^{12}$: $(3.3\pm1.0)\times 10^{6} \times\varepsilon_{K^0_{L}} $}   \\
\hline
$ J/\psi\rightarrow  \omega K^{\ast +}K^- \rightarrow \omega K^0_{S}\pi^+ K^- \rightarrow \pi^+\pi^-\pi^0 \pi^+\pi^-\pi^+ K^-$ & \tabincell{c}{$N_{J/\psi}=10^{10}$: $(4.6\pm0.7)\times 10^{4} \times\varepsilon_{K^0_{S}} $\\$N_{J/\psi}=10^{12}$: $(4.6\pm0.7)\times 10^{6} \times\varepsilon_{K^0_{S}} $}   \\
\hline
$ J/\psi\rightarrow  \omega K^{\ast +} K^- \rightarrow  \pi^+\pi^-\pi^0 K^0_{L}\pi^+ K^- $ & \tabincell{c}{$N_{J/\psi}=10^{10}$: $(6.0\pm0.9)\times 10^{4} \times\varepsilon_{K^0_{L}} $\\$N_{J/\psi}=10^{12}$: $(6.0\pm0.9)\times 10^{6} \times\varepsilon_{K^0_{L}} $}   \\
\hline
\end{tabular}
\end{center}
\end{table}

Here, we note that the interferences between $K^{\ast}$ meson and other $K^{\ast}$ mesons, such as $K^{\ast}(1680)$, $K^{\ast}_{2}(1430)$ and $K^{\ast}_{2}(1980)$, have non-negligible effects on the branching ratios of the $J/\psi\rightarrow  K^{\ast \pm} K^{\ast \mp} $, $J/\psi\rightarrow  K^{\ast \pm} K^\mp $, $J/\psi\rightarrow  K^{\ast \pm} K^{\mp}\pi^{0} $, $J/\psi\rightarrow  K^{\ast 0} K^{-}\pi^{+} $ and $J/\psi\rightarrow  \bar{K}^{\ast 0} K^{+}\pi^{-} $ decays~\cite{BESIII:2019apb,Ablikim:2010kd,BaBar:2007ceh,BaBar:2014uwz,BaBar:2017pkz}. Fortunately, the CP asymmetries in all the $K^{\ast\prime}\rightarrow K^0_{S,L}\pi$ decays can be derived
\begin{small}
\begin{align}
&\frac{{\mathcal B}\left(K^{\ast\prime}\rightarrow K^0_{S,L}\pi\right)-{\mathcal B}\left(\bar{K}^{\ast\prime}\rightarrow K^0_{S,L}\pi\right)}{{\mathcal B}\left(K^{\ast\prime}\rightarrow K^0_{S,L}\pi\right)+{\mathcal B}\left(\bar{K}^{\ast\prime}\rightarrow K^0_{S,L}\pi\right)}=\frac{{\mathcal B}\left(K^{\ast}\rightarrow K^0_{S,L}\pi\right)-{\mathcal B}\left(\bar{K}^{\ast}\rightarrow K^0_{S,L}\pi\right)}{{\mathcal B}\left(K^{\ast}\rightarrow K^0_{S,L}\pi\right)+{\mathcal B}\left(\bar{K}^{\ast}\rightarrow K^0_{S,L}\pi\right)},\label{Eq:interferekstar}
\end{align}
\end{small}
where $K^{\ast\prime}$ denotes all the $K^{\ast}$ meson except the $K^{\ast}(892)$ meson. Therefore the interferences between $K^{\ast}$ meson and other $K^{\ast}$ mesons have no effect on the CP violations in $K^{\ast}\rightarrow K^0_{S,L}\pi$ decays. Here, one can consider the possibility to sum over all the decays involving $K^{\ast}$ resonance in order to obtain a statistically significant signal of CP violation. From Eqs.(\ref{Eq:kstarkscpasy}), (\ref{Eq:kstarklcpasy}), (\ref{Eq:jpsikstarkslcp}) and (\ref{Eq:jpsieqkstarcp}), we can see that the resonance structures in the $K^{\ast \pm} K^\mp$ invariant mass spectrum of the $J/\psi\rightarrow  \eta^{\prime} K^{\ast \pm} K^\mp$ decays and the $K^{\ast 0} \bar{K}^{\ast 0}$ invariant mass spectrum of the $J/\psi\rightarrow \gamma K^{\ast 0} \bar{K}^{\ast 0}$ decay also have no effect on the CP violations in these decays~\cite{BES:1999zaa,BESIII:2018ede}.

Now, we proceed to study the $K_{S}^0-K_{L}^0$ asymmetries in the $K^{\ast}\rightarrow K^0_{S,L}\pi$ decays, which are defined as~\cite{Bigi:1994aw,Wang:2017ksn,CLEO:2007rhw}:
\begin{align}
&R\left(K^{\ast}\rightarrow K^0_{S,L}\pi\right)=\frac{{\mathcal B}\left(K^{\ast}\rightarrow K^0_{S}\pi\right)-{\mathcal B}\left(K^{\ast}\rightarrow K^0_{L}\pi\right)}{{\mathcal B}\left(K^{\ast}\rightarrow K^0_{S}\pi\right)+{\mathcal B}\left(K^{\ast}\rightarrow K^0_{L}\pi\right)},\label{Eq:ksklasykstarksl1}\\
&R\left(\bar{K}^{\ast}\rightarrow K^0_{S,L}\pi\right)=\frac{{\mathcal B}\left(\bar{K}^{\ast}\rightarrow K^0_{S}\pi\right)-{\mathcal B}\left(\bar{K}^{\ast}\rightarrow K^0_{L}\pi\right)}{{\mathcal B}\left(\bar{K}^{\ast}\rightarrow K^0_{S}\pi\right)+{\mathcal B}\left(\bar{K}^{\ast}\rightarrow K^0_{L}\pi\right)}.\label{Eq:ksklasykstarksl2}
\end{align}
Substituting Eqs.(\ref{Eq:kstarzksdened2}), (\ref{Eq:kstbarksdened1}), (\ref{Eq:kstarzkldecay}) and (\ref{Eq:kstarzbarkldecay}) into Eqs.(\ref{Eq:ksklasykstarksl1}) and (\ref{Eq:ksklasykstarksl2}) and assuming $z=0$, we can obtain
\begin{align}
&R\left(K^{\ast}\rightarrow K^0_{S,L}\pi\right)=\frac{1}{2}\frac{e^{-\Gamma_L t_0}- e^{-\Gamma_L t_1}}{e^{-\Gamma_S t_0}-e^{-\Gamma_S t_1}} \cdot\frac{ {\mathcal B}(K_L^0\rightarrow \pi^+\pi^-)}{{\mathcal B}(K_S^0\rightarrow \pi^+\pi^-)}+Re\left( t_{K^0_S-K_L^0}\cdot \frac{p-q}{p+q}\right),\label{Eq:ksklasykstarksl3}\\
&R\left(\bar{K}^{\ast}\rightarrow K^0_{S,L}\pi\right)=\frac{1}{2}\frac{e^{-\Gamma_L t_0}- e^{-\Gamma_L t_1}}{e^{-\Gamma_S t_0}-e^{-\Gamma_S t_1}} \cdot\frac{ {\mathcal B}(K_L^0\rightarrow \pi^+\pi^-)}{{\mathcal B}(K_S^0\rightarrow \pi^+\pi^-)}-Re\left( t_{K^0_S-K_L^0}\cdot \frac{p-q}{p+q}\right).\label{Eq:ksklasykstarksl4}
\end{align}
In the $J/\psi\rightarrow K^{\ast} f_{J/\psi}\rightarrow K^0_{S,L}\pi f_{J/\psi}$ and $J/\psi\rightarrow \bar{K}^{\ast}\bar{f}_{J/\psi}\rightarrow K^0_{S,L}\pi \bar{f}_{J/\psi}$ decays, the $K_{S}^0-K_{L}^0$ asymmetry can de defined as
\begin{align}
&R\left(J/\psi\rightarrow K^{\ast}f_{J/\psi}\rightarrow K^0_{S,L}\pi f_{J/\psi} \right)\nonumber\\
&=\frac{{\mathcal B}\left(J/\psi\rightarrow K^{\ast}f_{J/\psi}\rightarrow K^0_{S}\pi f_{J/\psi}\right)-{\mathcal B}\left(J/\psi\rightarrow K^{\ast}f_{J/\psi}\rightarrow K^0_{L}\pi f_{J/\psi}\right)}{{\mathcal B}\left(J/\psi\rightarrow K^{\ast}f_{J/\psi}\rightarrow K^0_{S}\pi f_{J/\psi}\right)+{\mathcal B}\left(J/\psi\rightarrow K^{\ast}f_{J/\psi}\rightarrow K^0_{L}\pi f_{J/\psi}\right)},\label{Eq:ksklasykstarksl5}\\
&R\left(J/\psi\rightarrow \bar{K}^{\ast} \bar{f}_{J/\psi}\rightarrow K^0_{S,L}\pi \bar{f}_{J/\psi} \right)\nonumber\\
&=\frac{{\mathcal B}\left(J/\psi\rightarrow \bar{K}^{\ast} \bar{f}_{J/\psi}\rightarrow K^0_{S}\pi \bar{f}_{J/\psi}\right)-{\mathcal B}\left(J/\psi\rightarrow \bar{K}^{\ast}\bar{f}_{J/\psi}\rightarrow K^0_{L}\pi \bar{f}_{J/\psi}\right)}{{\mathcal B}\left(J/\psi\rightarrow \bar{K}^{\ast} \bar{f}_{J/\psi}\rightarrow K^0_{S}\pi \bar{f}_{J/\psi}\right)+{\mathcal B}\left(J/\psi\rightarrow \bar{K}^{\ast} \bar{f}_{J/\psi}\rightarrow K^0_{L}\pi \bar{f}_{J/\psi}\right)},\label{Eq:ksklasykstarksl6}
\end{align}
Combining Eqs.(\ref{Eq:ksklasykstarksl1})-(\ref{Eq:ksklasykstarksl2}) with Eqs.(\ref{Eq:ksklasykstarksl5})-(\ref{Eq:ksklasykstarksl6}), we can obtain the following relations
\begin{align}
&R\left(J/\psi\rightarrow K^{\ast}f_{J/\psi}\rightarrow K^0_{S,L}\pi f_{J/\psi} \right)=R\left(K^{\ast}\rightarrow K^0_{S,L}\pi\right),\label{Eq:ksklasykstarksl7}\\
&R\left(J/\psi\rightarrow \bar{K}^{\ast} \bar{f}_{J/\psi}\rightarrow K^0_{S,L}\pi \bar{f}_{J/\psi} \right)=R\left(\bar{K}^{\ast}\rightarrow K^0_{S,L}\pi\right).\label{Eq:ksklasykstarksl8}
\end{align}
With Eqs.(\ref{Eq:ksklasykstarksl5})-(\ref{Eq:ksklasykstarksl8}), we can derive the numbers of the observed signal events on the $K_{S}^0-K_{L}^0$ asymmetry in $J/\psi$ decays
\begin{align}
&N_{K^0_{S}-K^0_{L}}^{K^{\ast}}=\left|R\left(K^{\ast}\rightarrow K^0_{S,L}\pi\right)\right| \cdot N_{J/\psi}\cdot \varepsilon_{K^0_{L}} \cdot {\mathcal B}\left(J/\psi\rightarrow K^{\ast}f_{J/\psi}\rightarrow K^0_{S}\pi f_{J/\psi}+ K^0_{L}\pi f_{J/\psi}\right) ,\label{Eq:numberksklasy1}\\
&N_{K^0_{S}-K^0_{L}}^{\bar{K}^{\ast}}=\left|R\left(\bar{K}^{\ast}\rightarrow K^0_{S,L}\pi\right)\right| \cdot N_{J/\psi}\cdot \varepsilon_{K^0_{L}}\cdot {\mathcal B}\left(J/\psi\rightarrow \bar{K}^{\ast} \bar{f}_{J/\psi}\rightarrow K^0_{S}\pi \bar{f}_{J/\psi}+ K^0_{L}\pi \bar{f}_{J/\psi}\right) ,\label{Eq:numberksklasy2}
\end{align}
Using the values of the parameters in Table~\ref{theparametervalue} and combining Eqs.(\ref{Eq:ksdefinition3}), (\ref{Eq:substitigamma}), (\ref{Eq:ksklasykstarksl3}) and (\ref{Eq:ksklasykstarksl4}), we can obtain
\begin{align}
&R\left(K^{\ast}\rightarrow K^0_{S,L}\pi\right)=\left(3.51\pm0.03\right)\times 10^{-3},\label{Eq:valueofksklasykskl1}\\
&R\left(\bar{K}^{\ast}\rightarrow K^0_{S,L}\pi\right)=\left(-3.45\pm0.03\right)\times 10^{-3}.\label{Eq:valueofksklasykskl2}
\end{align}
The branching ratios for the $J/\psi\rightarrow K^{\ast} f_{J/\psi}\rightarrow K^0_{S,L}\pi f_{J/\psi}$  and $J/\psi\rightarrow \bar{K}^{\ast} \bar{f}_{J/\psi}\rightarrow K^0_{S,L}\pi \bar{f}_{J/\psi}$ decays can be obtained directly from PDG~\cite{Zyla:2020zbs} and Table~\ref{totbranchrakstar}. With the values of the branching ratios for these decays and combining Eqs.(\ref{Eq:numberksklasy1})-(\ref{Eq:valueofksklasykskl2}), we calculate the numerical results of $N_{K^0_{S}-K^0_{L}}^{K^{\ast}}$ and $N_{K^0_{S}-K^0_{L}}^{\bar{K}^{\ast}}$, which are listed in Table~\ref{ksklasyextnumevent}, with $10^{10}$ $J/\psi$ event sample and $10^{12}$ $J/\psi$ event sample, respectively. Here, we also note that the detection efficiency $\varepsilon_{K^0_{L}}$ is  at the level of $ 10^{-3}$ at BESIII~\cite{BESIII:2021yam,BES:2007hqc}, so the $K_{S}^0-K_{L}^0$ asymmetry can be observed in the decays with $N_{K^0_{S}-K^0_{L}}^{K^{\ast}}\left(\text{or }N_{K^0_{S}-K^0_{L}}^{\bar{K}^{\ast}}\right)$ is larger than $10^{4} \times\varepsilon_{K^0_{L}}$.
\begin{table}[t]
\newcommand{\tabincell}[2]{\begin{tabular}{@{}#1@{}}#2\end{tabular}}
\begin{center}
\caption{\label{ksklasyextnumevent} \small The expected numbers of the observed signal events on the $K_{S}^0-K_{L}^0$ asymmetry in $J/\psi$ decays.}
\vspace{0.1cm}
\doublerulesep 0.8pt \tabcolsep 0.18in
\scriptsize
\begin{tabular}{c|c}
\hline
the decay channel & $N_{K^0_{S}-K^0_{L}}^{K^{\ast}}\left(N_{K^0_{S}-K^0_{L}}^{\bar{K}^{\ast}}\right)$ \\
\hline
$J/\psi\rightarrow  K^{\ast 0} \bar{K}^{\ast 0}\rightarrow K^0_{S,L}\pi^0 K^-\pi^+  $ & \tabincell{c}{$N_{J/\psi}=10^{10}$: $(1.8\pm0.5)\times 10^{3} \times\varepsilon_{K^0_{L}} $\\$N_{J/\psi}=10^{12}$: $(1.8\pm0.5)\times 10^{5} \times\varepsilon_{K^0_{L}}$}  \\
\hline
$J/\psi\rightarrow  K^{\ast 0} \bar{K}^{\ast 0}\rightarrow  K^+\pi^- K^0_{S,L}\pi^0 $ & \tabincell{c}{$N_{J/\psi}=10^{10}$: $(1.8\pm0.5)\times 10^{3} \times\varepsilon_{K^0_{L}} $\\$N_{J/\psi}=10^{12}$: $(1.8\pm0.5)\times 10^{5} \times\varepsilon_{K^0_{L}}$}  \\
\hline
$ J/\psi\rightarrow  K^{\ast +} K^{\ast -}\rightarrow K^0_{S,L}\pi^+ K^-\pi^0 $ & \tabincell{c}{$N_{J/\psi}=10^{10}$: $(7.8^{-3.1}_{+1.7})\times 10^{3} \times\varepsilon_{K^0_{L}} $\\$N_{J/\psi}=10^{12}$: $(7.8^{-3.1}_{+1.7})\times 10^{5} \times\varepsilon_{K^0_{L}} $}  \\
\hline
$ J/\psi\rightarrow  K^{\ast +} K^{\ast -}\rightarrow K^+\pi^0 K^0_{S,L}\pi^-$ & \tabincell{c}{$N_{J/\psi}=10^{10}$: $(7.7^{-3.1}_{+1.7})\times 10^{3} \times\varepsilon_{K^0_{L}} $\\$N_{J/\psi}=10^{12}$: $(7.7^{-3.1}_{+1.7})\times 10^{5} \times\varepsilon_{K^0_{L}} $}  \\
\hline
$ J/\psi\rightarrow \eta K^{\ast 0} \bar{K}^{\ast 0}\rightarrow \gamma \gamma K^0_{S,L}\pi^0 K^-\pi^+$ & \tabincell{c}{$N_{J/\psi}=10^{10}$: $(3.5\pm0.8)\times 10^{3} \times\varepsilon_{K^0_{L}} $\\$N_{J/\psi}=10^{12}$: $(3.5\pm0.8)\times 10^{5} \times\varepsilon_{K^0_{L}} $}  \\
\hline
$ J/\psi\rightarrow \eta K^{\ast 0} \bar{K}^{\ast 0}\rightarrow \gamma \gamma K^+\pi^- K^0_{S,L}\pi^0$ & \tabincell{c}{$N_{J/\psi}=10^{10}$: $(3.5\pm0.8)\times 10^{3} \times\varepsilon_{K^0_{L}} $\\$N_{J/\psi}=10^{12}$: $(3.5\pm0.8)\times 10^{5} \times\varepsilon_{K^0_{L}} $}  \\
\hline
$ J/\psi\rightarrow \gamma K^{\ast 0} \bar{K}^{\ast 0}\rightarrow \gamma K^0_{S,L}\pi^0 K^-\pi^+ $ & \tabincell{c}{$N_{J/\psi}=10^{10}$: $(3.1\pm1.0)\times 10^{4} \times\varepsilon_{K^0_{L}} $\\$N_{J/\psi}=10^{12}$: $(3.1\pm1.0)\times 10^{6} \times\varepsilon_{K^0_{L}} $}  \\
\hline
$ J/\psi\rightarrow \gamma K^{\ast 0} \bar{K}^{\ast 0}\rightarrow \gamma  K^+\pi^- K^0_{S,L}\pi^0$ & \tabincell{c}{$N_{J/\psi}=10^{10}$: $(3.1\pm1.0)\times 10^{4} \times\varepsilon_{K^0_{L}} $\\$N_{J/\psi}=10^{12}$: $(3.1\pm1.0)\times 10^{6} \times\varepsilon_{K^0_{L}} $}  \\
\hline
$ J/\psi\rightarrow  \eta^{\prime} K^{\ast +}K^- \rightarrow  \pi^+\pi^-\eta K^0_{S,L}\pi^+ K^-\rightarrow\pi^+\pi^-\gamma\gamma K^0_{S,L}\pi^+ K^-$ & \tabincell{c}{$N_{J/\psi}=10^{10}$: $(2.9\pm0.3)\times 10^{3} \times\varepsilon_{K^0_{L}} $\\$N_{J/\psi}=10^{12}$: $(2.9\pm0.3)\times 10^{5} \times\varepsilon_{K^0_{L}} $}  \\
\hline
$J/\psi\rightarrow  \eta^{\prime}  K^{\ast -} K^+\rightarrow \pi^+\pi^-\eta K^0_{S,L}\pi^- K^+   \rightarrow \pi^+\pi^-\gamma\gamma K^0_{S,L}\pi^- K^+ $ & \tabincell{c}{$N_{J/\psi}=10^{10}$: $(2.9\pm0.3)\times 10^{3} \times\varepsilon_{K^0_{L}} $\\$N_{J/\psi}=10^{12}$: $(2.9\pm0.3)\times 10^{5} \times\varepsilon_{K^0_{L}} $}  \\
\hline
$ J/\psi\rightarrow  K^{\ast +}K^- \rightarrow  K^0_{S,L}\pi^+ K^- $ & \tabincell{c}{$N_{J/\psi}=10^{10}$:  $(7.0^{-1.2}_{+0.9})\times 10^{4} \times\varepsilon_{K^0_{L}} $\\$N_{J/\psi}=10^{12}$: $(7.0^{-1.2}_{+0.9})\times 10^{6} \times\varepsilon_{K^0_{L}} $}  \\
\hline
$  J/\psi\rightarrow K^{\ast -}K^+  \rightarrow  K^0_{S,L}\pi^-K^+$ & \tabincell{c}{$N_{J/\psi}=10^{10}$: $(6.9^{-1.2}_{+0.9})\times 10^{4} \times\varepsilon_{K^0_{L}} $\\$N_{J/\psi}=10^{12}$: $(6.9^{-1.2}_{+0.9})\times 10^{6} \times\varepsilon_{K^0_{L}} $}  \\
\hline
$J/\psi\rightarrow  K^{\ast 0} K^-\pi^+\rightarrow K^0_{S,L}\pi^0 K^-\pi^+ $ & \tabincell{c}{$N_{J/\psi}=10^{10}$: $(4.5\pm0.9)\times 10^{4} \times\varepsilon_{K^0_{L}} $\\$N_{J/\psi}=10^{12}$: $(4.5\pm0.9)\times 10^{6} \times\varepsilon_{K^0_{L}} $}  \\
\hline
$J/\psi\rightarrow \bar{K}^{\ast 0} K^+\pi^-\rightarrow K^0_{S,L}\pi^0 K^+\pi^-$ & \tabincell{c}{$N_{J/\psi}=10^{10}$: $(4.4\pm0.9)\times 10^{4} \times\varepsilon_{K^0_{L}} $\\$N_{J/\psi}=10^{12}$: $(4.4\pm0.9)\times 10^{6} \times\varepsilon_{K^0_{L}} $}  \\
\hline
$J/\psi\rightarrow  K^{\ast +} K^-\pi^0\rightarrow K^0_{S,L}\pi^+ K^-\pi^0  $ & \tabincell{c}{$N_{J/\psi}=10^{10}$: $(3.5\pm1.1)\times 10^{4} \times\varepsilon_{K^0_{L}} $\\$N_{J/\psi}=10^{12}$: $(3.5\pm1.1)\times 10^{6} \times\varepsilon_{K^0_{L}} $}  \\
\hline
$J/\psi\rightarrow K^{\ast -} K^+\pi^0\rightarrow K^0_{S,L}\pi^- K^+\pi^0  $ & \tabincell{c}{$N_{J/\psi}=10^{10}$: $(3.5\pm1.1)\times 10^{4} \times\varepsilon_{K^0_{L}} $\\$N_{J/\psi}=10^{12}$: $(3.5\pm1.1)\times 10^{6} \times\varepsilon_{K^0_{L}} $}  \\
\hline
$J/\psi\rightarrow  \omega K^{\ast +}K^- \rightarrow \pi^+\pi^-\pi^0  K^0_{S,L}\pi^+ K^- $ & \tabincell{c}{$N_{J/\psi}=10^{10}$: $(6.4\pm0.9)\times 10^{4} \times\varepsilon_{K^0_{L}} $\\$N_{J/\psi}=10^{12}$: $(6.4\pm0.9)\times 10^{6} \times\varepsilon_{K^0_{L}} $}  \\
\hline
$ J/\psi\rightarrow \omega K^{\ast -} K^+ \rightarrow \pi^+\pi^-\pi^0  K^0_{S,L}\pi^- K^+ $ & \tabincell{c}{$N_{J/\psi}=10^{10}$: $(6.3\pm0.9)\times 10^{4} \times\varepsilon_{K^0_{L}} $\\$N_{J/\psi}=10^{12}$: $(6.3\pm0.9)\times 10^{6} \times\varepsilon_{K^0_{L}} $}  \\
\hline
\end{tabular}
\end{center}
\end{table}

The $K^{\ast}\rightarrow K^0_{S,L}\pi$ decays can also be used to study the CPT violation. We define the following observables, which are related to the CPT violation parameter $z$
\begin{align}
&{\mathcal A}_{CPT}^{m}\left(K^{\ast}\rightarrow K^0_{S,L}\pi\right)=\frac{{\mathcal A}_{K^0_{S}}^{-}-{\mathcal A}_{K^0_{L}}^{-}}{{\mathcal A}_{K^0_{S}}^{+}+{\mathcal A}_{K^0_{L}}^{+}},\label{Eq:cptkstarkslasy1}\\
&{\mathcal A}_{CPT}^{p}\left(K^{\ast}\rightarrow K^0_{S,L}\pi\right)=\frac{{\mathcal A}_{K^0_{S}}^{+}-{\mathcal A}_{K^0_{L}}^{+}}{{\mathcal A}_{K^0_{S}}^{+}+{\mathcal A}_{K^0_{L}}^{+}},\label{Eq:cptkstarkslasy2}
\end{align}
where
\begin{align}
&{\mathcal A}_{K^0_{S,L}}^{\pm}={\mathcal B}\left(K^{\ast}\rightarrow K^0_{S,L}\pi\right)\pm{\mathcal B}\left(\bar{K}^{\ast}\rightarrow K^0_{S,L}\pi\right).\label{Eq:cptkstarkslasy3}
\end{align}
Substituting Eqs.(\ref{Eq:kstarzksdened2}), (\ref{Eq:kstbarksdened1}), (\ref{Eq:kstarzkldecay}) and (\ref{Eq:kstarzbarkldecay}) into Eqs.(\ref{Eq:cptkstarkslasy1}) and (\ref{Eq:cptkstarkslasy2}), we can easily find
\begin{align}
&{\mathcal A}_{CPT}^{m}\left(K^{\ast}\rightarrow K^0_{S,L}\pi\right)=-Re(z)+Re\left( t_{K^0_S-K_L^0}\cdot \left(\frac{p-q}{p+q}+\frac{z}{2}\right)\right),\label{Eq:cptkstarkslasy4}\\
&{\mathcal A}_{CPT}^{p}\left(\bar{K}^{\ast}\rightarrow K^0_{S,L}\pi\right)=-Re(z)\cdot\frac{\left|q\right|^2-\left|p\right|^2}{\left|q\right|^2+\left|p\right|^2}+\frac{1}{2}\frac{e^{-\Gamma_L t_0}- e^{-\Gamma_L t_1}}{e^{-\Gamma_S t_0}-e^{-\Gamma_S t_1}} \cdot\frac{ {\mathcal B}(K_L^0\rightarrow \pi^+\pi^-)}{{\mathcal B}(K_S^0\rightarrow \pi^+\pi^-)}\nonumber\\
&~~~~~~~~~~~~~~~~~~~~~~~~~~~~~~~~~~~~~+ Re\left(\left(\frac{\left|q\right|^2-\left|p\right|^2}{\left|q\right|^2+\left|p\right|^2}+i Im(z)\right) t_{K^0_S-K_L^0}\cdot \left(\frac{p-q}{p+q}+\frac{z}{2}\right)\right),\label{Eq:cptkstarkslasy5}
\end{align}
obviously,  ${\mathcal A}_{CPT}^{m,p}\left(K^{\ast}\rightarrow K^0_{S,L}\pi\right)$ contains the terms $t_{K^0_S-K_L^0}\cdot\frac{p-q }{p+q }$ and $(\left|q\right|^2-\left|p\right|^2)/(\left|q\right|^2+\left|p\right|^2)$, which are independent from the CPT violation parameter $z$, so the precise calculations of $t_{K^0_S-K_L^0}\cdot\frac{p-q }{p+q }$ and $(\left|q\right|^2-\left|p\right|^2)/(\left|q\right|^2+\left|p\right|^2)$, which are the functions of the parameters $m_L-m_S$, $\Gamma_L$, $\Gamma_S$, $p$, $q$, $t_0$ and $t_1$, is crucial to constraint the CPT violation parameter $z$ in the $K^{\ast}\rightarrow K^0_{S,L}\pi$ decays.

In the $J/\psi$ decays involving the $K^{\ast}$ meson in the final states, we define the CPT asymmetry observables
\begin{align}
&{\mathcal A}_{CPT}^{m}\left(J/\psi\rightarrow  K^{\ast} f_{J/\psi}\right)=\frac{{\mathcal A}_{J/\psi K^0_{S}}^{-}-{\mathcal A}_{J/\psi K^0_{L}}^{-}}{{\mathcal A}_{J/\psi K^0_{S}}^{+}+{\mathcal A}_{J/\psi K^0_{L}}^{+}},\label{Eq:cptkstarkslasy6}\\
&{\mathcal A}_{CPT}^{p}\left(J/\psi\rightarrow  K^{\ast} f_{J/\psi}\right)=\frac{{\mathcal A}_{J/\psi K^0_{S}}^{+}-{\mathcal A}_{J/\psi K^0_{L}}^{+}}{{\mathcal A}_{J/\psi K^0_{S}}^{+}+{\mathcal A}_{J/\psi K^0_{L}}^{+}},\label{Eq:cptkstarkslasy7}
\end{align}
where
\begin{align}
&{\mathcal A}_{J/\psi K^0_{S}}^{\pm}={\mathcal B}\left(J/\psi\rightarrow  K^{\ast} f_{J/\psi}\rightarrow  K^0_{S}\pi f_{J/\psi}\right)\pm{\mathcal B}\left(J/\psi\rightarrow \bar{K}^{\ast} \bar{f}_{J/\psi}\rightarrow K^0_{S}\pi\bar{f}_{J/\psi} \right),\label{Eq:cptkstarkslasy8}\\
&{\mathcal A}_{J/\psi K^0_{L}}^{\pm}={\mathcal B}\left(J/\psi\rightarrow  K^{\ast} f_{J/\psi}\rightarrow  K^0_{L}\pi f_{J/\psi}\right)\pm{\mathcal B}\left(J/\psi\rightarrow \bar{K}^{\ast} \bar{f}_{J/\psi}\rightarrow K^0_{L}\pi\bar{f}_{J/\psi} \right).\label{Eq:cptkstarkslasy9}
\end{align}
Obviously, according to Eqs.(\ref{Eq:cptkstarkslasy1})-(\ref{Eq:cptkstarkslasy3}) and Eqs.(\ref{Eq:cptkstarkslasy6})-(\ref{Eq:cptkstarkslasy9}), we can obtain
\begin{align}
&{\mathcal A}_{CPT}^{m}\left(J/\psi\rightarrow  K^{\ast} f_{J/\psi}\right)={\mathcal A}_{CPT}^{m}\left(K^{\ast}\rightarrow K^0_{S,L}\pi\right),\label{Eq:cptkstarkslasy10}\\
&{\mathcal A}_{CPT}^{p}\left(J/\psi\rightarrow  K^{\ast} f_{J/\psi}\right)={\mathcal A}_{CPT}^{p}\left(K^{\ast}\rightarrow K^0_{S,L}\pi\right).\label{Eq:cptkstarkslasy11}
\end{align}
With the values of the parameters in Table~\ref{theparametervalue} and combining Eqs.(\ref{Eq:ksdefinition3}), (\ref{Eq:substitigamma}), (\ref{Eq:cptkstarkslasy4}) and (\ref{Eq:cptkstarkslasy5}), we can obtain
\begin{align}
&Re\left( t_{K^0_S-K_L^0}\cdot \left(\frac{p-q}{p+q}\right)\right)=\left(3.48\pm0.03\right)\times 10^{-3},\label{Eq:cptkstarkslasy12}\\
&\frac{1}{2}\frac{e^{-\Gamma_L t_0}- e^{-\Gamma_L t_1}}{e^{-\Gamma_S t_0}-e^{-\Gamma_S t_1}} \cdot\frac{ {\mathcal B}(K_L^0\rightarrow \pi^+\pi^-)}{{\mathcal B}(K_S^0\rightarrow \pi^+\pi^-)}+\frac{\left|q\right|^2-\left|p\right|^2}{\left|q\right|^2+\left|p\right|^2}\cdot Re\left( t_{K^0_S-K_L^0}\cdot \frac{p-q}{p+q}\right)=\left(1.54\pm0.03\right)\times 10^{-5},\label{Eq:cptkstarkslasy13}
\end{align}
which accuracy can reach $10^{-4}$ and $10^{-6}$, respectively. Combining Eqs.(\ref{Eq:cptkstarkslasy4})-(\ref{Eq:cptkstarkslasy5}) with Eqs.(\ref{Eq:cptkstarkslasy10})-(\ref{Eq:cptkstarkslasy13}), we can obtain the numerical results of ${\mathcal A}_{CPT}^{m,p}\left(J/\psi\rightarrow  K^{\ast} f_{J/\psi}\right)$ under the assumption of $z=0$
\begin{align}
&{\mathcal A}_{CPT}^{m}\left(J/\psi\rightarrow  K^{\ast} f_{J/\psi}\right)_{z=0}=\left(3.48\pm0.03\right)\times 10^{-3},\label{Eq:cptkstarkslasy14}\\
&{\mathcal A}_{CPT}^{p}\left(J/\psi\rightarrow  K^{\ast} f_{J/\psi}\right)_{z=0}=\left(1.54\pm0.03\right)\times 10^{-5}.\label{Eq:cptkstarkslasy15}
\end{align}
From Eqs.(\ref{Eq:cptkstarkslasy6})-(\ref{Eq:cptkstarkslasy7}) and taking into account only the statistical errors, the errors of ${\mathcal A}_{CPT}^{m,p}\left(J/\psi\right.\rightarrow  K^{\ast}\left. f_{J/\psi}\right)$ can be derived
\begin{align}
&\Delta\left({\mathcal A}_{CPT}^{m,p}\left(J/\psi\rightarrow  K^{\ast} f_{J/\psi}\right)\right)\approx\frac{1}{\sqrt{N_{J/\psi}\cdot\left({\mathcal A}_{J/\psi K^0_{S}}^{+}+{\mathcal A}_{J/\psi K^0_{L}}^{+}\right)\cdot \varepsilon_{K^0_{L}}}}.\label{Eq:cptkstarkslasy16}
\end{align}
From PDG~\cite{Zyla:2020zbs} and Table~\ref{totbranchrakstar}, the branching ratios for $J/\psi\rightarrow K^{\ast} f_{J/\psi}\rightarrow K^0_{S,L}\pi f_{J/\psi}$ can reach the level of $10^{-3}$, so the errors $\Delta\left({\mathcal A}_{CPT}^{m,p}\left(J/\psi\rightarrow  K^{\ast} f_{J/\psi}\right)\right)$ can arrive at the level of $ 10^{-4}$, if we assume the selection efficiency is $10^{-3}$ and the total number of $J/\psi$ events is $10^{12}$. According to these result, Eq.(\ref{Eq:cptkstarkslasy4}) and Eq.(\ref{Eq:cptkstarkslasy12}), we can obtain that the sensitivity for the measurement of the CPT violation parameter $z$ is expected to be at the level of  $10^{-4}$ with ${\mathcal A}_{CPT}^{m}\left(J/\psi\rightarrow  K^{\ast} f_{J/\psi}\right)$ at BESIII. Because there exist suppression effect on the CPT violation parameter $z$ in Eq.(\ref{Eq:cptkstarkslasy5}), the observable ${\mathcal A}_{CPT}^{p}\left(J/\psi\rightarrow  K^{\ast} f_{J/\psi}\right)$ is insensitive to the measurement of $z$ with $10^{12}$ $J/\psi$ event sample. Currently, the best result for $Re(z)$ is $-(5.2\pm5.0)\times 10^{-4}$ which is obtained from a combined fit, including KLOE ~\cite{KLOE-2:2018yif,DAmbrosio:2006hes} and CPLEAR~\cite{CPLEAR:1998dvs}, by the Particle Data Group~\cite{Zyla:2020zbs}, so the measurement of the CPT violation parameter $z$ is expected to be competitive with the current best result with $10^{12}$ $J/\psi$ event sample at BESIII. Here, we note that the sensitivity of the measurement of the CPT violation parameter $z$ depends on  measured precision of the parameters $m_L-m_S$, $\Gamma_L$, $\Gamma_S$, $p$ and $q$ and the consistence between the values of $t_0$ and $t_1$ and the event selection criteria in experiment.

In addition, we can define the following observable
\begin{align}
&{\mathcal A}_{CPT}\left(K^{\ast}\rightarrow K^0_{L}\pi\right)=\frac{4|p|^2 \cdot {\mathcal B}\left(K^{\ast}\rightarrow K^0_{L}\pi\right)-4|q|^2 \cdot {\mathcal B}\left(\bar{K}^{\ast}\rightarrow K^0_{L}\pi\right)}{4|p|^2 \cdot {\mathcal B}\left(K^{\ast}\rightarrow K^0_{L}\pi\right)+4|q|^2 \cdot {\mathcal B}\left(\bar{K}^{\ast}\rightarrow K^0_{L}\pi\right)},\label{Eq:cptkstarkslasy17}
\end{align}
substituting Eqs.(\ref{Eq:kstarzkldecay}), (\ref{Eq:kstarzbarkldecay}) into Eq.(\ref{Eq:cptkstarkslasy17}), we can find
\begin{align}
&{\mathcal A}_{CPT}\left(K^{\ast}\rightarrow K^0_{L}\pi\right)=\frac{Re(z)}{1+\frac{|z|^2}{4}},\label{Eq:cptkstarkslasy18}
\end{align}
which is sensitive to the CPT violation parameter $Re(z)$. However, the sensitivity of the measurement of the CPT violation parameter $Re(z)$ with the observable ${\mathcal A}_{CPT}\left(K^{\ast}\rightarrow K^0_{L}\pi\right)$ depends on the measured precisions of the parameters $p$ and $q$.

In conclusion, we discuss the CP asymmetry, $K_S^0-K_L^0$ asymmetry and CPT asymmetry in $K^{\ast}\rightarrow   K^0_{S,L}\pi$ decays at BESIII. The CP asymmetries in the $K^{\ast}\rightarrow   K^0_{S,L}\pi$ decays are dominated by $K^0-\bar{K}^0$ mixing. We calculate the numerical results of the CP asymmetries in the $J/\psi$ decays involving the $K^{\ast}$ meson in the final states
\begin{align}
{\mathcal A}_{CP}\left(J/\psi\rightarrow K^{\ast}f_{J/\psi}\rightarrow K^0_{S}\pi f_{J/\psi}\right)=\left(3.64\pm0.04\right)\times 10^{-3},\label{Eq:conacpkstarkspi}\\
{\mathcal A}_{CP}\left(J/\psi\rightarrow K^{\ast} f_{J/\psi}\rightarrow K^0_{L}\pi f_{J/\psi}\right)=\left(-3.32\pm0.04\right)\times 10^{-3}. \label{Eq:conacpkstarklpi}
\end{align}
We estimate the expected numbers of the observed signal events on the CP violations in $J/\psi$ decays with $K^{\ast}$ meson in the final states based on $10^{10}$ $J/\psi$ event sample and $10^{12}$ $J/\psi$ event sample in BESIII experiment, respectively. We find that the BESIII experiment may be able to make a significant measurement of the CP violation in these decays with $10^{10}$ $J/\psi$ events, which has been accumulated in four runs in 2009, 2012, 2018 and 2019~\cite{BESIII:2020nme,Yuan:2021yks}.

The $K_S^0-K_L^0$ asymmetries in $J/\psi$ decays with $K^{\ast}$ meson in the final states are also studied. The numerical results of the $K_S^0-K_L^0$ asymmetries can be obtain
\begin{align}
&R\left(J/\psi\rightarrow K^{\ast}f_{J/\psi}\rightarrow K^0_{S,L}\pi f_{J/\psi} \right)=\left(3.51\pm0.03\right)\times 10^{-3},\label{Eq:convalofksklasykskl1}\\
&R\left(J/\psi\rightarrow \bar{K}^{\ast} \bar{f}_{J/\psi}\rightarrow K^0_{S,L}\pi \bar{f}_{J/\psi} \right)=\left(-3.45\pm0.03\right)\times 10^{-3}.\label{Eq:convalofksklasykskl2}
\end{align}
Together with these results and the branching ratios for the $J/\psi$ decays, we calculate the expected numbers of the observed signal events on the $K_S^0-K_L^0$ asymmetries in the case of $10^{10}$ $J/\psi$ event sample and $10^{12}$ $J/\psi$ event sample in BESIII experiment, respectively. If we assume that the detection efficiency $\varepsilon_{K^0_{L}}$ is  at the level of $ 10^{-3}$ at BESIII, the $K_{S}^0-K_{L}^0$ asymmetry can be observed in the decays with the expected number of the observed signal events on the $K_S^0-K_L^0$ asymmetry is larger than $10^{4} \times\varepsilon_{K^0_{L}}$.

We investigate the possibility to constraint the CPT violation parameter $z$ in $J/\psi$ decays with $K^{\ast}$ meson in the final states at BESIII. We discuss the sensitivity for the measurement of the CPT violation parameter $z$ under the assumption that the selection efficiency is $10^{-3}$ and the total number of $J/\psi$ events is $10^{12}$. We find that the sensitivity for the measurement of the CPT violation parameter $z$ is expected to be at the level of $10^{-4}$ with $10^{12}$ $J/\psi$ event sample at BESIII, which could be competitive with the current best result. The sensitivity of the measurement of the CPT violation parameter $z$ depend on  measured precision of the parameters $m_L-m_S$, $\Gamma_L$, $\Gamma_S$, $p$ and $q$ and the consistence between the values of $t_0$ and $t_1$ and the event selection criteria in experiment.

The work was supported by the National Natural Science Foundation of China (Contract Nos. 12175088, 11805077) and the Key Scientific Research Projects of Colleges and Universities in Henan Province (Contract No. 18A140029).

\end{document}